\documentclass[referee,sn-nature]{sn-jnl}%

\usepackage{graphicx}
\usepackage{multirow}
\usepackage{amsmath,amssymb,amsfonts}
\usepackage{amsthm}
\usepackage{mathrsfs}
\usepackage[title]{appendix}
\usepackage{xcolor}
\usepackage{booktabs}
\usepackage{longtable}
\usepackage[section]{placeins}
\newcommand{\edfig}[1]{Extended Data Fig.~\ref{#1}}
\newcommand{\edtab}[1]{Extended Data Table~\ref{#1}}

\usepackage{todonotes}

\definecolor{llgray}{gray}{0.95}

\usepackage[draft]{changes}  
\definechangesauthor[color=brown]{K}
\definechangesauthor[color=blue]{AM}
\definechangesauthor[color=blue]{H}
\usepackage{xspace}
\newcommand{\mas}{\,mas\xspace}
\def\mJup{\,\textrm{M}$_\textrm{Jup}$\xspace}

\def\mSun{\,\textrm{M}$_\odot$\xspace}

\def\masyr{\,$\textrm{mas}\,\textrm{yr}^{-1}$\xspace}
\def\host{HR\,8799\xspace}

\def\si{Supplementary Information\xspace}
\def\au{{\,au}\xspace}

\begin{document}
\title{A fifth companion in the HR 8799 system revealed by Gaia } 
\author*[1]{\fnm{A.-M.} \sur{ Lagrange}}\email{anne-marie.lagrange@obspm.fr}
\author[2]{\fnm{K.} \sur{Go\'zdziewski }}\email{k.gozdziewski@umk.pl}
\author[1]{\fnm{F.} \sur{Kiefer }}\email{flavien.kiefer@obspm.fr}
\author[3]{\fnm{H.} \sur{Beust }}\email{Herve.Beust@univ-grenoble-alpes.fr}
\author[4]{\fnm{P.} \sur{ Rubini}}\email{pascal.rubini@pixyl.ai}
\author[5,6]{\fnm{ A.} \sur{ Zurlo}}\email{alice.zurlo@mail.udp.cl}
\author[1]{\fnm{ P.} \sur{ Thebault}}\email{philippe.thebault@obspm.fr}
\author[1]{\fnm{ A.} \sur{ Boccaletti}}\email{anthony.boccaletti@obspm.fr}
\affil*[1]{LIRA, CNRS, Observatoire de Paris, CNRS, Universite PSL, 5 Place Jules Janssen, 92190 Meudon, France}
\affil[2]{Institute of Astronomy, Faculty of Physics, Astronomy and Informatics, Nicolaus Copernicus University, Grudzia\c{a}dzka 5, PL-87-100 Toru\'n, Poland}
\affil[3]{Univ. Grenoble Alpes, CNRS, IPAG, 38000 Grenoble, France}
\affil[4]{Pixyl, 5 av. du Grand Sablon, 38700 La Tronche, France}
\affil[5]{Instituto de Estudios Astrof\'isicos, Facultad de Ingenier\'ia y Ciencias, Universidad Diego Portales, Av. Ej\'ercito Libertador 441, Santiago, Chile}
\affil[6]{Millennium Nucleus on Young Exoplanets and their Moons (YEMS)}


\abstract{
\host{} is the benchmark young multi-planet system, hosting four directly imaged giant planets orbiting between a warm inner belt and a cold outer debris ring. However, the few-au region has remained observationally inaccessible, lying beyond the reach of conventional high-contrast imaging and challenging for radial-velocity techniques due to stellar pulsations. Here we combine Gaia absolute astrometry and Hipparcos-Gaia proper-motion anomalies with radial velocities and high-contrast imaging to reveal an additional close-in substellar companion. The allowed solutions span $\sim$0.3-6\,au, with posteriors peaking at $\sim$2-3\,au and with masses $\sim$10-14\,\mJup. Dynamical modelling shows that such a companion with a low eccentricity can coexist with the long-lived resonant chain of the four outer planets, and sculpt the inner edge of the warm debris belt, linking the planet architecture to the debris-disk morphology. Our results close a long-standing detection gap in HR 8799 and demonstrate the power of Gaia astrometry to uncover massive planets at intermediate separations that may remain undetectable with radial velocities or direct imaging alone.
}

\maketitle

\section*{Introduction}

HR 8799 is the archetype of directly imaged planetary systems, uniquely revealing multiple massive planets locked in a resonant configuration at wide separations. Four giant planets were discovered through high-contrast imaging at projected separations between 14 and 68\,au \citep{Marois2008,Marois2011}. The system is young ($<30$\,Myr, possibly $10$–$23$\,Myr; \citealp{Sepulveda2022}) and dynamically remarkable: the four planets form a near-coplanar chain with adjacent period ratios close to 2:1, consistent with long-term stability and resonant locking \citep[e.g.,][]{Fabrycky2010,Gozdziewski2014,Wang2018,Zurlo2022}.  

HR~8799 also hosts a structured debris disk. Early modelling of the system's  spectral energy distribution revealed a warm ($\sim150$\,K) inner belt located interior to planet~\host{}e, with an inner edge at $\simeq 5$-$6$\,au, as well as a cold ($\simeq 45$\,K) outer tenuous component beyond the outermost planet \host{}b, and a halo of cold small grains \citep{Su2009}. Recent JWST observations provide direct evidence of the warm inner belt \citep{Boccaletti2024}, while millimetre imaging resolves a broad outer ring extending from $\sim100$–150\,au to $\sim300$\,au \citep{Booth2016,Wilner2018,Faramaz2021,Matra2025}.  

The known planets alone cannot fully account for the observed debris morphology. It was recognized that a planet beyond \host{}b was needed to explain the inner edge of the outer disk \citep{Booth2016,Read2018,Wilner2018,Gailer2019}. In addition, the  inner edge of the warm belt at $\sim5$–6\,au lies well inside the orbit of planet~e, suggesting dynamical sculpting by an additional inner companion. 

Establishing whether HR~8799 hosts such an inner body is also essential to determine whether the currently known four-planet chain represents the complete planetary architecture. However, direct imaging searches remain inconclusive at a few au. A tentative Keck/NIRC2 signal suggests a $4$--$7\,M_{\rm Jup}$ source at $\sim$4--5\,au \citep{Thompson2023}, while deeper VLT/SPHERE analyses do not confirm this detection \citep{Wahhaj2021,Dallant2023}. Detection limits via direct imaging  are, however, very sensitive to data processing details, especially at short separations, on the assumed age and considered cooling tracks. The radial velocity technique, very successful in detecting planets around solar to late-type stars, is not sensitive to planetary-mass companions around strongly pulsating $\gamma $  Dor early-type stars such as HR8799 \citep{Lagrange2013}. Absolute astrometry provides a complementary route to probe the intermediate-separation regime, offering sensitivity to massive companions that may evade both imaging and radial velocities.

\section*{An inner substellar companion in the \host{} system}
We use GaiaPMEX, a tool recently developed \citep[][see Methods]{Kiefer2025a} to model the Gaia astrometric renormalized unit weight errors (RUWE), and, when available, Gaia-Hipparcos Proper Motion anomalies based on Gaia DR3 release \citep[hereafter PMa, ][]{Brandt2021,Kervella2022,2026A&A...705A.238V}. Both diagnostics show significant excesses  ($3.3\sigma$ and $2.7\sigma$, respectively) relative to a single-star solution.  

A stellar origin to these excesses is discarded (see Methods), leaving a companion as a natural explanation. Interpreted as a single-companion signature, the excess in Gaia astrometry (hereafter called Gaia excess) favours a close-in ($\sim$0.05-20au) substellar object, and the PMa yields a broader ($\sim$0.05-200au) semi-major-axis range (Fig.~\ref{fig:fig1}). Given their masses and semi-major axes, the planets bcde are located well below the Gaia excess solutions  (Fig.~\ref{fig:fig1}) and do not contribute to the Gaia excess. Hence an additional companion is needed to explain the Gaia excess. On the contrary, planets bcde, and in particular, planet e, contribute to the PMa.  

Because the PMa is sensitive to the known planets bcde, we forward-modeled the contribution of planets bcde using the published dynamical solution of \citep{Zurlo2022} (Model~1 of their Table~7) propagated to the Gaia reference epoch (2016.0), and subtract this contribution from both the PMa signal (see Methods). We performed an analogous correction for the Gaia excess, even though the impact is expected to be small. The new PMa map (Fig.~\ref{fig:fig2}, top right) is no longer indicative of the presence of a companion, while the new Gaia map (Fig.~\ref{fig:fig2}, top left) 
is very similar to that of Fig.~\ref{fig:fig1}. This shows that the outer planets account for the measured PMa, while the Gaia excess cannot be explained by the known four-planet architecture alone. Analyzing the Gaia excess and PMa jointly leads to a combined map that provides the possible (mass, sma) of this close-in object (Fig.~\ref{fig:fig2}, bottom).

Finally, we combine Gaia astrometry with PMa, archival radial velocities  (\edtab{tab:EX-RV}), and high-contrast imaging limits from \citep{Dallant2023} in a joint MCMC analysis (see Methods and \edtab{tab:EX-MCMCpriors}). Radial velocities constrain short-period stellar companions, while direct imaging is expected to constrain the widest-orbit solutions. The resulting posterior confines the companion to $\sim$0.3-6\,au and with a substellar mass (Fig.~\ref{fig:fig3}; see also \edfig{fig:EX_CORNER}). The highest posterior density is found for $a\simeq2$-3\,au and $m\simeq10$-15\,\mJup (peak at 11.5\,\mJup), i.e., in the planetary-mass regime. Two additional families of solutions appear in the brown-dwarf regime, centered near (0.8au, 40\,\mJup) and (0.5au, 80\,\mJup), at substantially lower posterior density. The cumulative posterior probabilities for these three solution families are 78.5$\% $, 7$\% $, and 14.5$\% $, respectively.

\section*{Architecture and stability of the HR8799 bcdef system}
We tested whether the range of masses and orbits inferred for the innermost companion is compatible with the long-term dynamical stability of the outer planets over at least the system age ($\simeq$30\,Myr). The bcde subsystem has been commonly interpreted as a coplanar near-resonant 8e:4d:2c:1b chain that can remain stable for hundreds of Myr \citep[][and references therein]{Zurlo2022}. Because this architecture reproduces the bcde relative astrometry without significant residuals, we require viable five-planet solutions to preserve the outer resonant chain.

Adding an inner giant planet invalidates the published four-planet orbital solution due to significant barycenter shifts. It  apparently disrupts the resonant chain and triggers rapid dynamical instability (see SI). Therefore, introducing such a companion requires a self-consistent re-optimization of the osculating elements for the entire system. 
 
To construct five-planet resonant architectures that fully preserve the resonant bcde configuration described by \citep{Zurlo2022}, and remain stable over long timescales (up to $\simeq$360\,Myr), we used the Migration Constrained Optimization Algorithm \citep[MCOA,][]{Gozdziewski2014}. Our approach is described in the Methods, with implementation details provided in the Supplementary Information.

We first considered a putative planet f within the highest posterior-density region, and found viable solutions for the bcdef system by initializing from Model~1 of \citep{Zurlo2022}. \edfig{fig:EX-migration}  illustrates a representative MCOA outcome for a $m_{\rm f}\sim$ 11\mJup planet with $a_{\rm f}$ initiated at $\sim 3$\,au, in which resonant capture produces Laplace MMR chain configurations (\edtab{tab:EX-models}; see also Supplementary Information). Figure~\ref{fig:orb_evol11} shows the orbital evolution in the case of a $m_{\rm f}$=11.5\mJup and $a_{\rm f}\simeq$ 2.3au (Model~115), which reproduces the bcde orbital evolution of Model~1 in \citep{Zurlo2022} (see also \edfig{fig:EX-migration}). 

To assess the stability domain of planet~\host{}f while preserving  the bcde resonance chain, we  computed dynamical maps around representative solutions using a chaos fast indicator (see Methods). The stable domain for planet~\host{}f with a mass of $\simeq10$-15\,\mJup{}, within the dominant GaiaPMEX posterior mode spans several au in $a_{\rm f}$ and extends to $e_{\rm f}\sim0.15$-0.4 (Fig.~\ref{fig:dynmaps}), with the exact extent depending on the mass of planet~f. However, one should note that the  actual extent and shape of the stable domain also depend in a complex way on the initial conditions in the multi-parameter space. The bcde resonance regions remain confined to narrow stability-equilibrium islands with typical widths of $\simeq 0.3$\,au, comparable to  Model~1 of \citep{Zurlo2022}.  Planet~f weakly interacts with the resonant subsystem, as indicated by the preservation of the bcde Laplace MMR: the adiabatic tuning allows to recover this resonance (\si{}). 
 
The weak dynamical coupling between \host{}f and the outer resonant chain also explains why four-planet astrometric fits have not yet revealed \host{}f: its signature is small and largely averages out over the available $\sim$30\,yr baseline, which spans $\simeq 10$ revolutions of planet~f at $\simeq 2.5$\,au. 

Finally, we considered  more massive, closer-in companions, such as a brown dwarf of $\sim 40$\,\mJup at $\sim 0.8$\,au, or even $\simeq 80$\,\mJup at $\simeq 0.5$\,au, corresponding to the secondary GaiaPMEX solution families (see \si{}). The period ratio for these close-in objects is larger than that of the objects at  $\sim 2-3$\,au by more than an order of magnitude, and the relative astrometry signal averages as well (see above). We verified that the outer bcde Laplace mean-motion–resonance chain can also coexist with such companions (\edfig{fig:EX-els1444}).

We conclude that a 10–15\mJup{} planet at $\sim$3\,au is fully consistent with the current stability constraints implied by the bcde astrometry-based architecture and is further favored by its location within the resonant stability island in the dynamical maps. Shorter periods, more massive companions cannot be excluded by dynamical stability considerations alone.

\section*{Impact of HR8799f on the inner disk component}
We asked whether the inner companion is compatible with the warm inner debris disk. The 24\,$\mu$m excess measured by Spitzer and attributed to dust between $\sim$6 and 15\,au \citep{Su2009} cannot be due to the companion itself: BHAC15 model atmospheres \citep{Baraffe2015} predict  $\sim$0.08\,mJy (resp. 0.6 \,mJy) at 24$\,\mu$m for a 10 (resp. 80)\mJup{} object at 30\,Myr and 40\,pc, well below the observed $\sim$30\,mJy excess. Hence, the companion does not significantly contribute to the Spitzer flux. 
We then study its impact as a dynamical perturber of the dust-producing planetesimal belt.

Following classic Hill or Wisdom resonance overlap criterion \citep{Wisdom1980},  a Jovian planet with $m_{\rm f}=10$\mJup  and semi-major-axis $\simeq 3$\au would be expected to clear the disk out to a critical radius  $a_{\rm crit}$ of $\simeq 4.4$\au, while material beyond $a_{\rm crit}$ would remain largely preserved. We recall that \citep{Zurlo2022} showed that planet~e confines stable material to MMR-protected zones extending inward, below $\simeq 10$\,au. 

To connect these approximate constraints to our orbit-mass  solutions for planet f, we performed direct, long-term $N$-body debris-disk simulations for rigorously stable five-planet solutions (Models 10, 11, 115, and 14 in \edtab{tab:EX-models}) with $m_{\rm f} = 10-14$\mJup and orbiting at $2.3$–$3.3$\,au.  The disk is assumed to be coplanar with the planets, in the Laplace plane of the system. We initiated an annulus of massless, low-eccentricity particles between 2.5 and 10\,au, encompassing the orbit of planet~f and extending to the outer boundary of the disk. The system was integrated over the 35\,Myr. Details are provided in \si{}. Over $\sim$35\,Myr, the disk relaxes into a long-lived configuration, leaving a substantial population of planetesimals   concentrated between $4$–$8$\,au. An illustration is provided in Figure~\ref{fig:disk11}  in the case of a  11.5\mJup planet. We  conclude that a close-in, 10-14\mJup planet can naturally account for the inner depletion of material while still allowing for the presence of the inner belt sandwiched between planets e and f. 

 Finally, we examined the closer-in, higher-mass companion solutions suggested by the GaiaPMEX analysis. Using the Hill/Wisdom resonance-overlap criterion, we found that the corresponding clearing radius is $\lesssim 0.5$\,au for a $m_{\rm f}=80$\,\mJup object at $a_{\rm f}=0.3$\,au. Such a companion would therefore not disrupt the inner belt. Still, it would fail to account for the belt’s inner truncation, implying that additional companion(s) would be required to carve the inner edge.

\section*{Conclusions}
Absolute astrometry reveals an additional close-in companion in the \host{} system, interior to the four directly imaged giants. Combining Gaia astrometric excess and Hipparcos-Gaia proper-motion anomalies with radial velocities, high-contrast imaging limits and dynamical constraints confines the companion to substellar masses at separations of $\sim$0.3-6\,au, with the highest posterior density at $\sim $2-3\,au and  $m\sim$ 10-15\,\mJup. 
 Viable five-planet solutions involving a planet f in this range preserve the long-lived resonant architecture of the outer bcde chain consistent with prior models based on the relative astrometry, and can account for an inner void of material up to about 5 au, and a warm inner debris belt, linking the inner planet architecture to the observed disk morphology. 

With five massive companions totaling a mass of at least 30\mJup, \host{} becomes an extreme laboratory for the dynamics of tightly packed multi-giant systems. The inferred configuration strengthens the case that resonant capture and migration shaped the orbits of the outer planets, while an additional inner companion can simultaneously remain dynamically decoupled from the outer Laplace-type chain and sculpt the inner belt. The system  provides a rare opportunity to test models of multi-planet resonant assembly together with debris-disk confinement in a young system.

Future observations are needed to refine the orbit and mass of \host{} f and test the dynamical picture. A direct detection would yield relative astrometry and atmospheric constraints; depending on its true separation and luminosity, \host{} f may be accessible to VLTI/GRAVITY or JWST/AMI \citep{Desdoigts2025}. In any case, it is  an attractive target for ELT first-light instruments such as MICADO \citep{Baudoz2019} and METIS. Continued astrometric monitoring, together with improved high-contrast imaging and dynamical modeling, should enable a fully self-consistent reconstruction of this benchmark five-giant system.

The uncovered planet belongs to a system that is already known to host outer planets. A few inner companions have been identified via absolute astrometry around AB Pic and HD\,106906, both of which host very wide ($>100$\au) planets \citep{Lagrange2025,Bifani2023}), while a planet was recently found by combining RVs and absolute astrometry in the HD 206893 system, already known to host a closer-in brown dwarf  \citep{Grandjean2020}. \host{}, however, remains the only imaged system in which the giant planets form a resonant chain. Gaia astrometry can play a key role in completing planetary inventories and in connecting planet populations to debris-disk structures.

\host{}f also exemplifies the capability of Gaia to uncover planets at a few au that are not within reach of radial velocity techniques because of star's properties (early type stars, high rotation velocity) or pole-on orbital plane inclinations, and where direct imaging is challenged by contrast at inner working angles. This foreshadows Gaia’s broader impact on exoplanet demographics: many apparently well-characterized systems are likely to host unseen giant planets at intermediate separations, with important consequences for the population of cold Jupiters and the structure of debris disks. 
\clearpage

\begin{figure*}
\centerline{
\hbox{
\hbox{\includegraphics[width=0.5\textwidth]{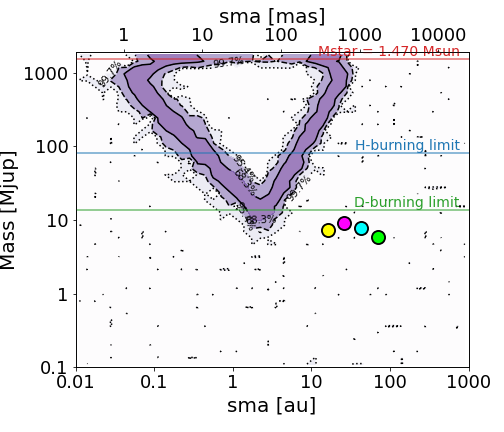}}
\hbox{\includegraphics[width=0.5\textwidth]{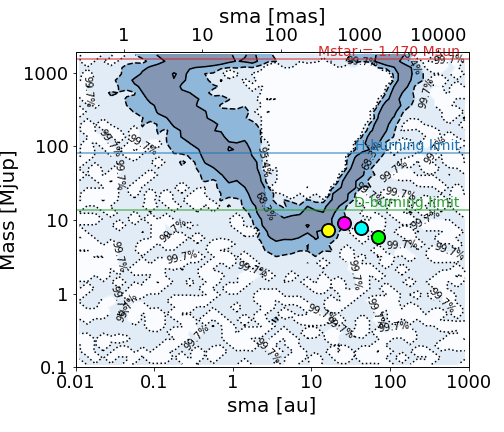}}
}
}
\caption{
 Mass-semi-major-axis solutions obtained by interpreting the Gaia astrometric excess (left) and the Hipparcos-Gaia proper-motion anomaly (right) as a single-companion signal. The Gaia excess robustly indicates an additional close-in companion, whereas the PMa is potentially contaminated by the known outer planets (see text and Methods). Planets bcde \citep{Zurlo2022} are indicated by plain round symbols with  colors respectively green, cyan, pink and yellow.  
}
\label{fig:fig1}
\end{figure*}

\begin{figure*}
\hbox{
\hbox{\includegraphics[width=0.5\textwidth]{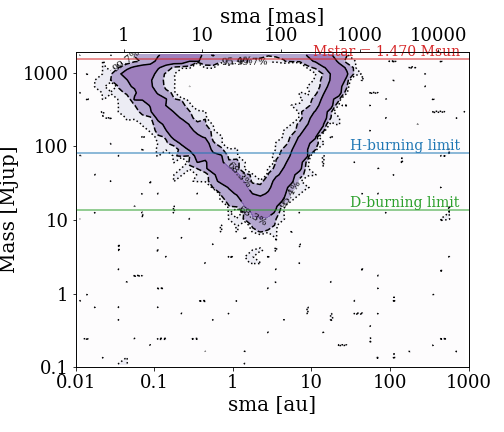}}
\hbox{\includegraphics[width=0.5\textwidth]{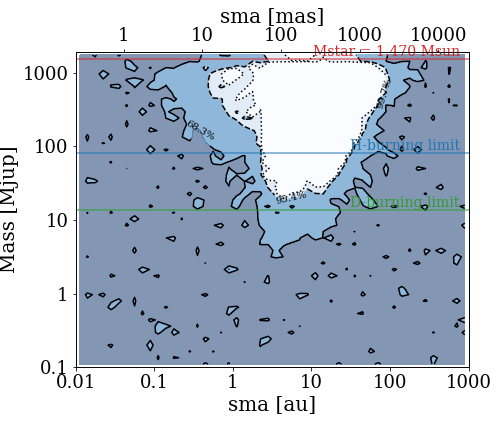}}
}
\hbox{
\hbox{\includegraphics[width=0.5\textwidth]{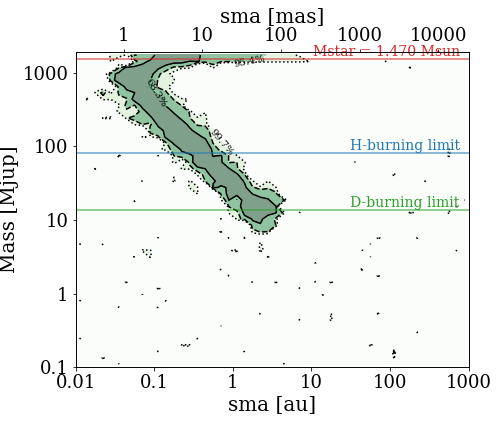}}
}
\caption{ 
Mass-semi-major-axis solutions when taking into account the planets bcde signals. Upper left: Gaia astrometric excess, which still requires an additional inner companion. Upper right: Hipparcos-Gaia PMa, which becomes consistent with the contribution from the known planets. Bottom: joint constraint from Gaia excess and PMa under the single-companion assumption. 
}
\label{fig:fig2}
\end{figure*}

\begin{figure*}
\vbox{
\centerline{\hbox{\includegraphics[width=1.\textwidth]{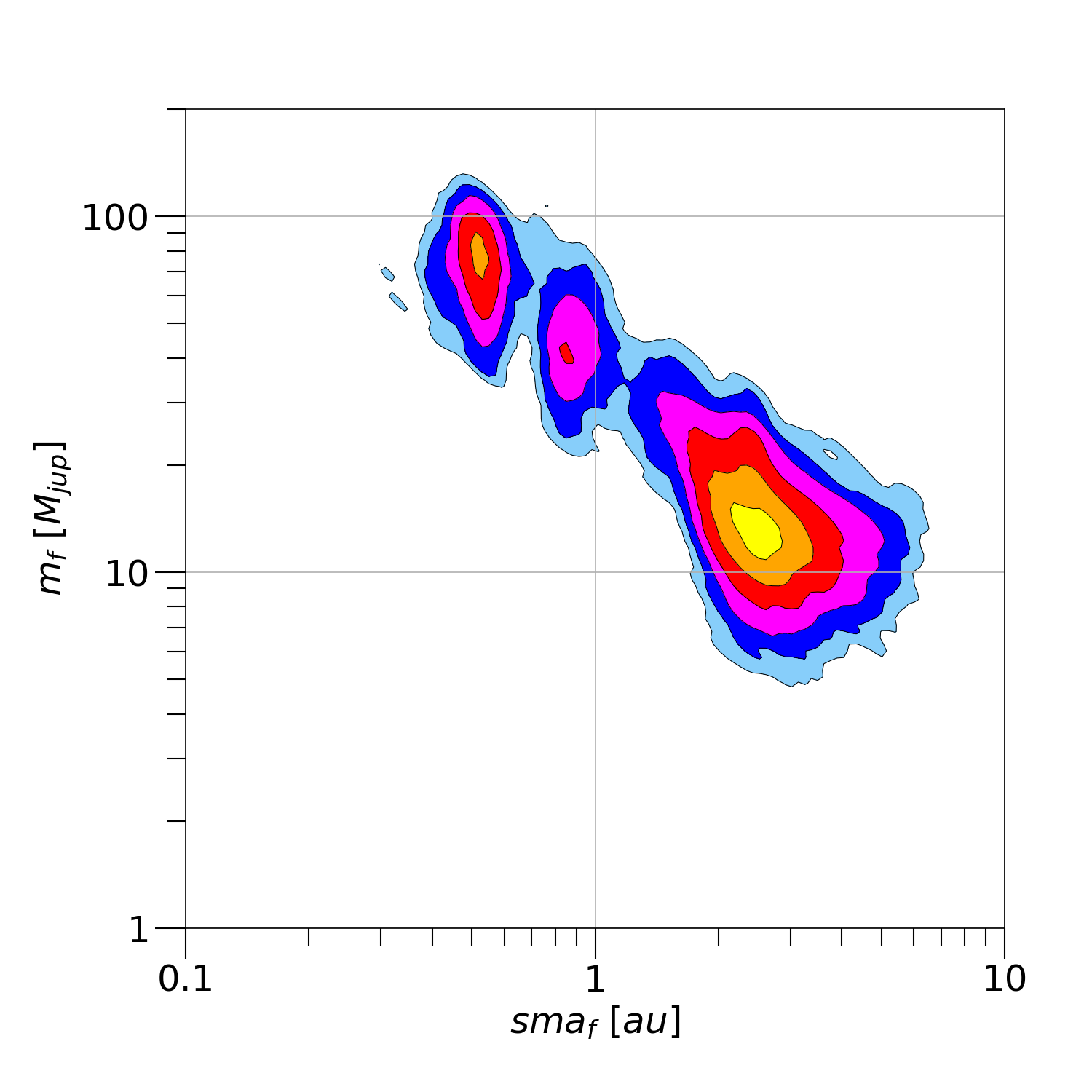}}}
}
\caption{
Constraints on the inner companion from the combination of absolute astrometry, radial velocities, and direct imaging.  Posterior from a joint fit combining Gaia astrometry, Hipparcos-Gaia PMa, radial velocities, and direct-imaging limits (Methods). The colors represent increasing likelihood, light blue: $3\sigma$,  blue: $2.5\sigma$, magenta: $2\sigma$, red: $1.5\sigma$, orange: $1\sigma$, yellow: $0.5\sigma$. 
}
\label{fig:fig3}
\end{figure*}

\begin{figure*}
\centerline{
\vbox{
\hbox{\includegraphics[width=0.8\textwidth]{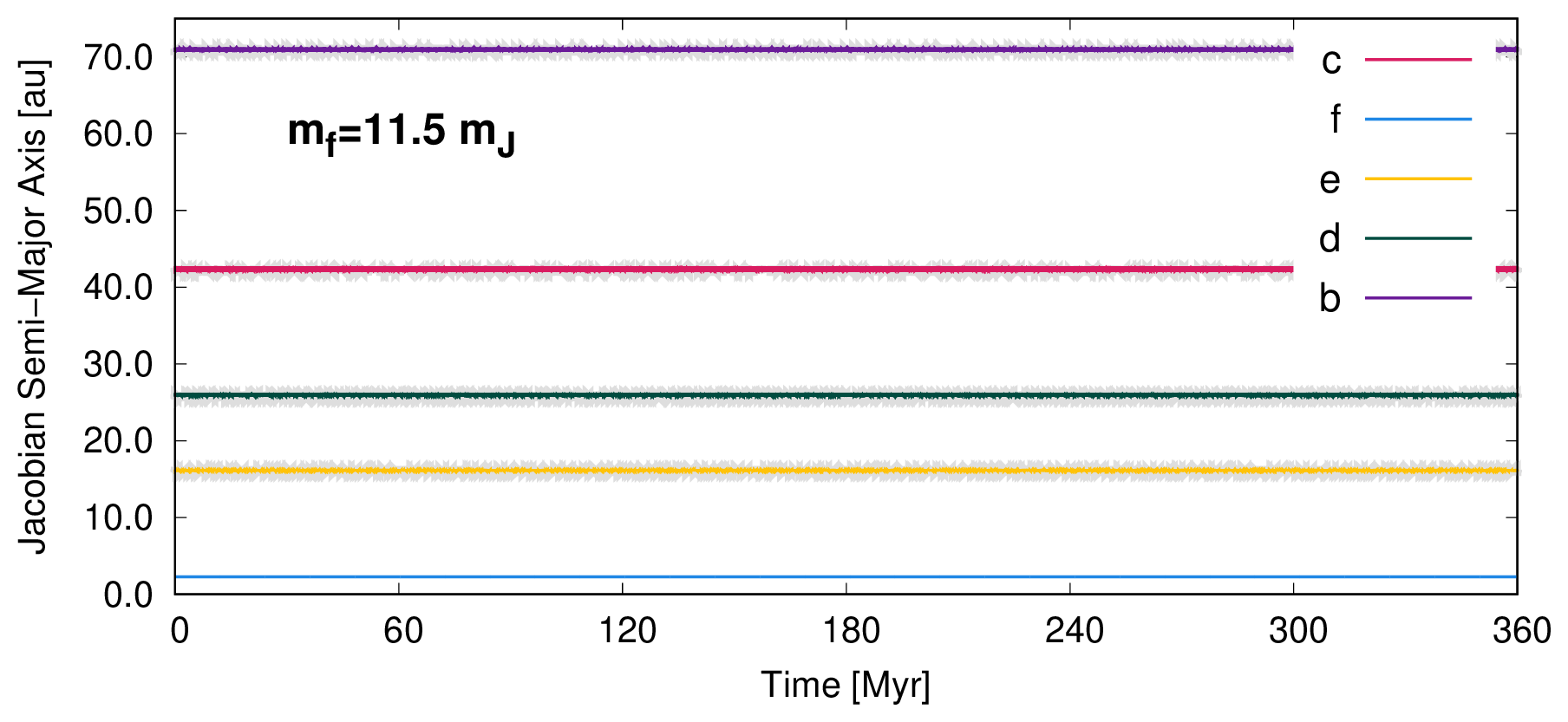}}
\hbox{\includegraphics[width=0.8\textwidth]{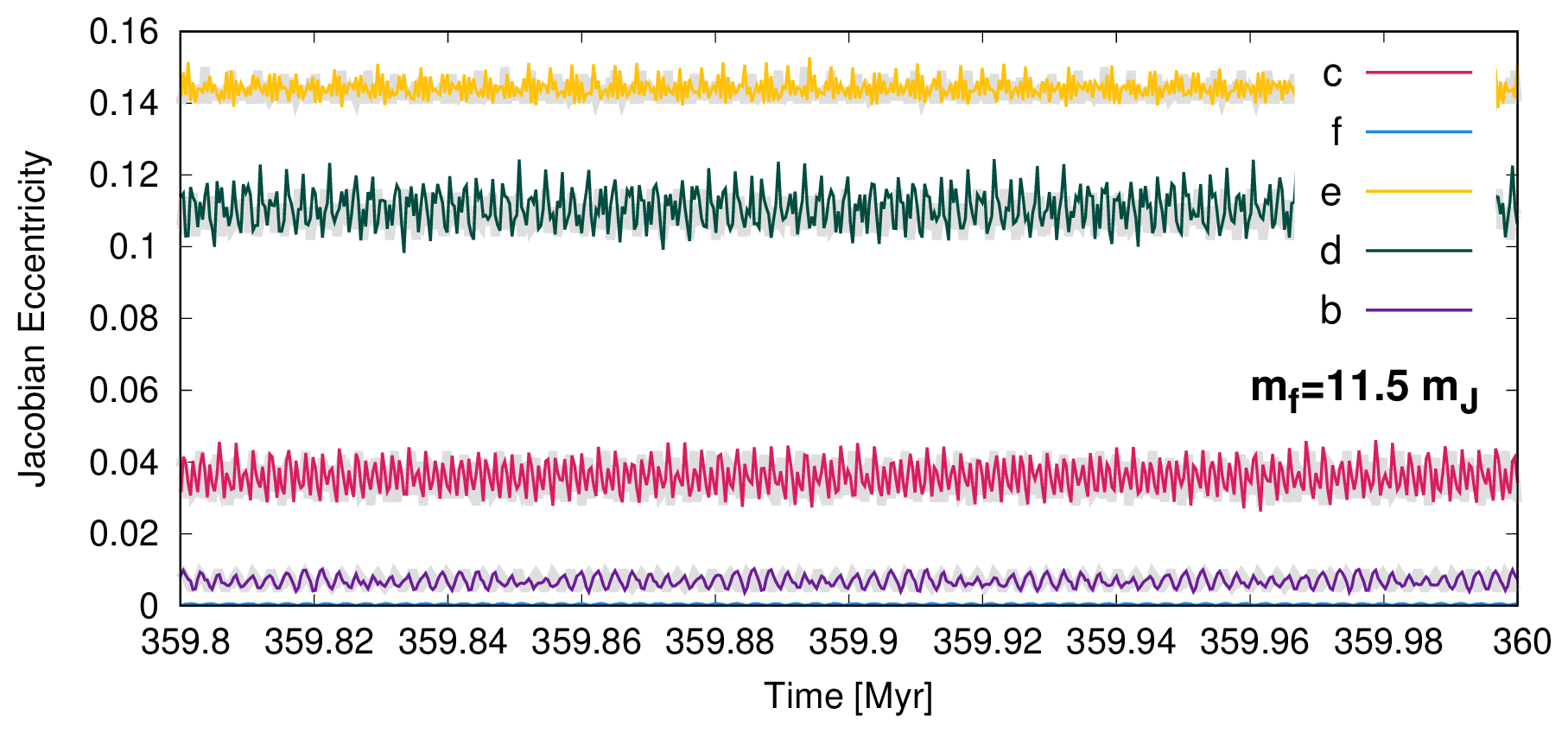}}
\hbox{\includegraphics[width=0.8\textwidth]{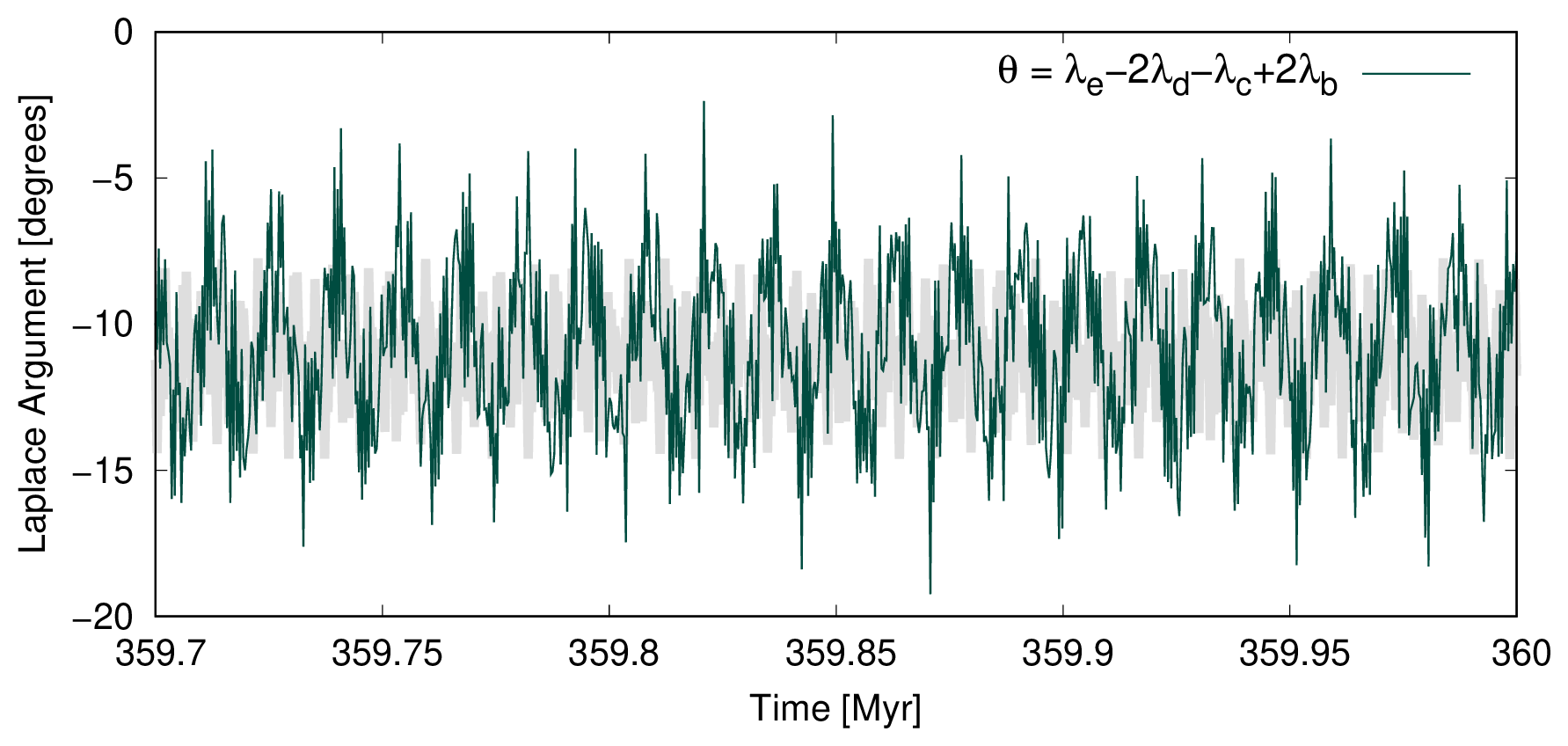}}
}
}
\caption{ {\bf
Evolution of the orbital elements in terms of Jacobi coordinates,  Model~11.5 listed in 
\edtab{tab:EX-models}
for a $11.46$\mJup planet orbiting at 2.3\,au.}  For comparison, shaded graphs indicate  the same elements in the exact Laplace MMR for four-planet Model~1 in \citep{Zurlo2022}.
{\em Top}: Semi-major axes over 360\,Myr (see Methods).
{\em Middle}: Eccentricity over the last 300\,Kyr of the integration interval.
{\em Bottom}: Evolution of the Laplace critical angle 
$\theta_{8:4:2:1} =  \lambda_{\rm e} - 2 \lambda_{\rm d} - \lambda_{\rm c} + 2\lambda_{\rm b}$ at the end of the integration interval (see SI). Sustained libration of this Laplace argument with low amplitude is a primary signature of a stable bcde configuration.
}
\label{fig:orb_evol11}
\end{figure*}

\begin{figure*}
\centerline{
\hbox{\includegraphics[width=0.48\textwidth]{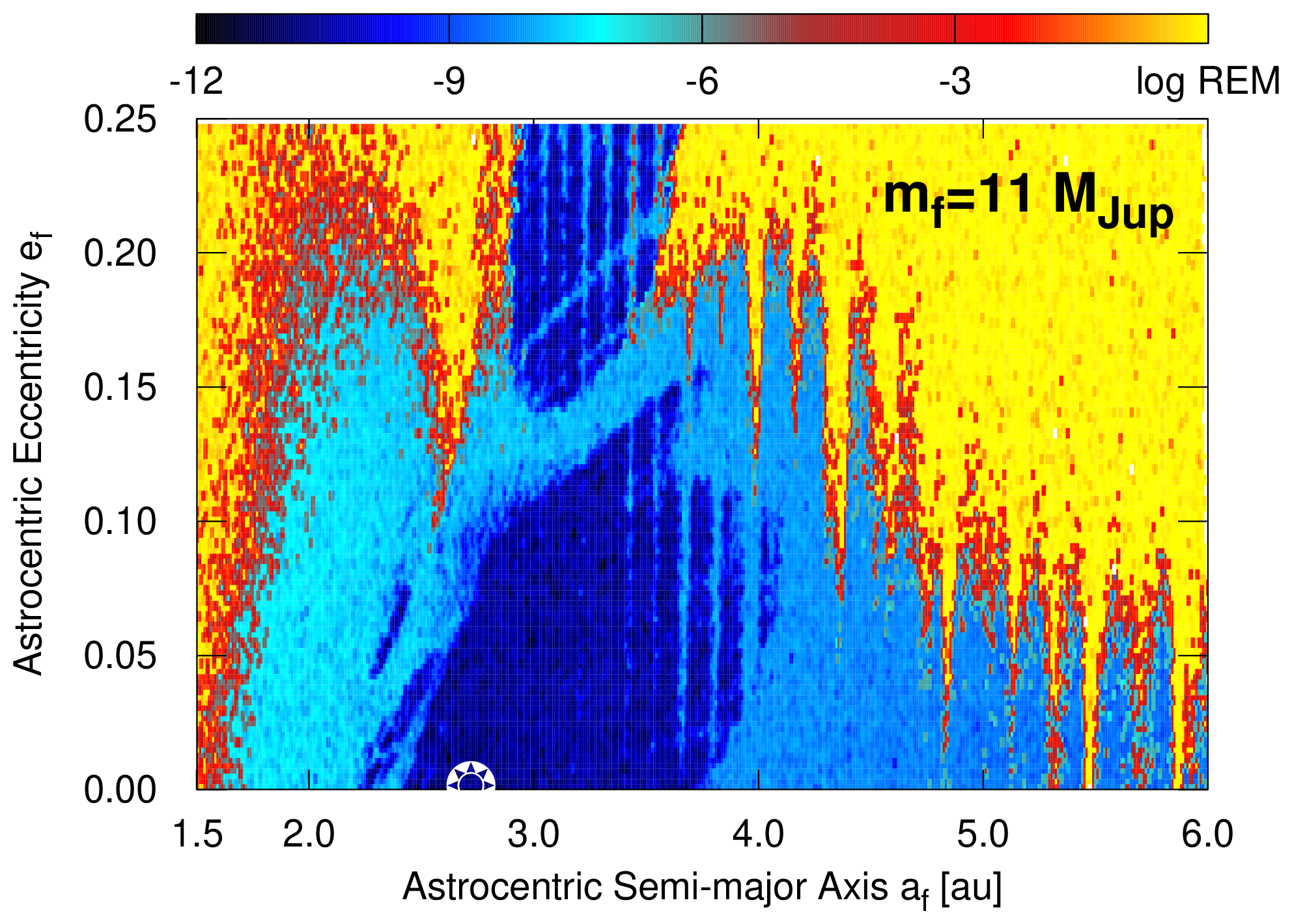}}
\hbox{\includegraphics[width=0.48\textwidth]{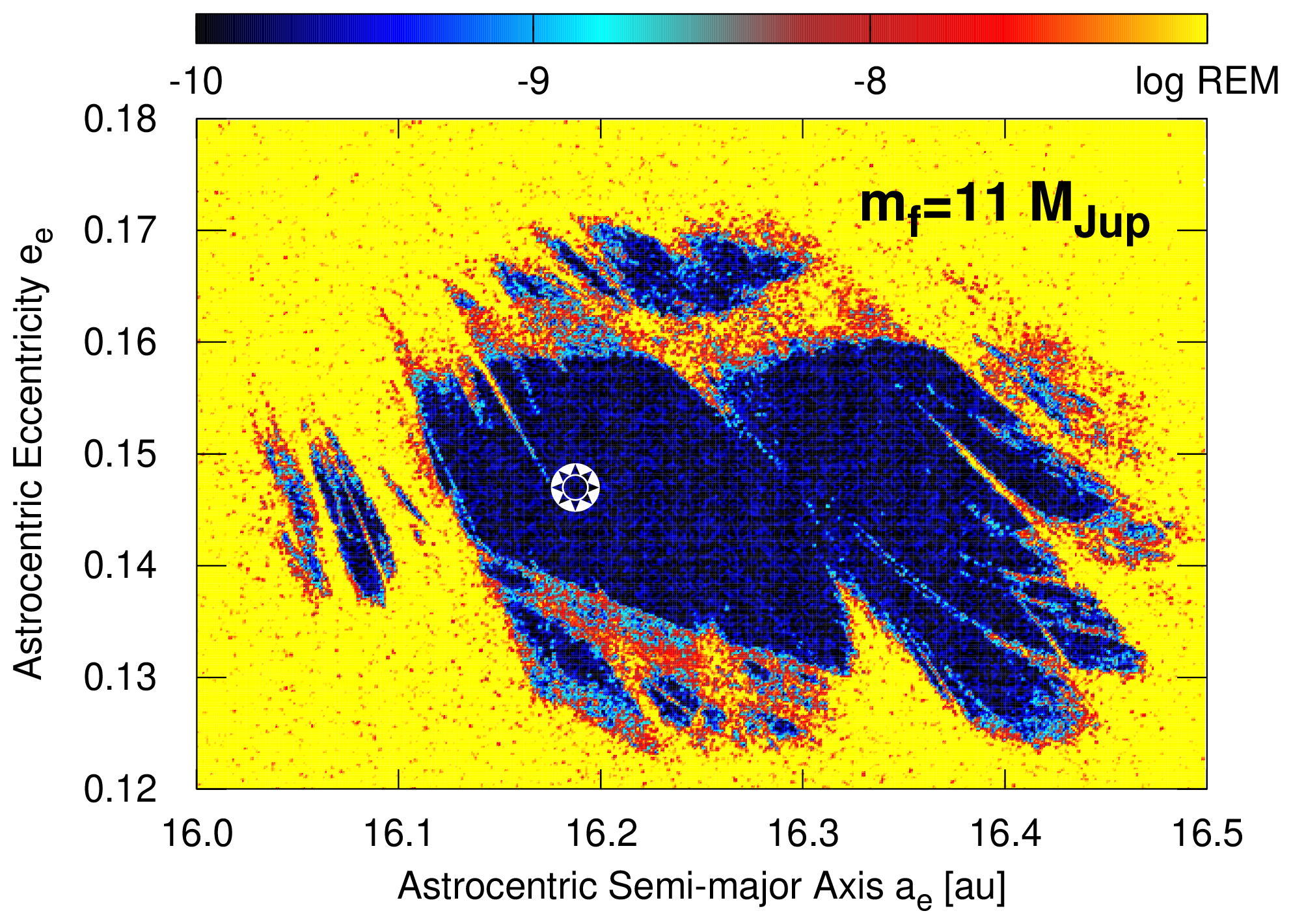}}
}
\centerline{
\hbox{\includegraphics[width=0.48\textwidth]{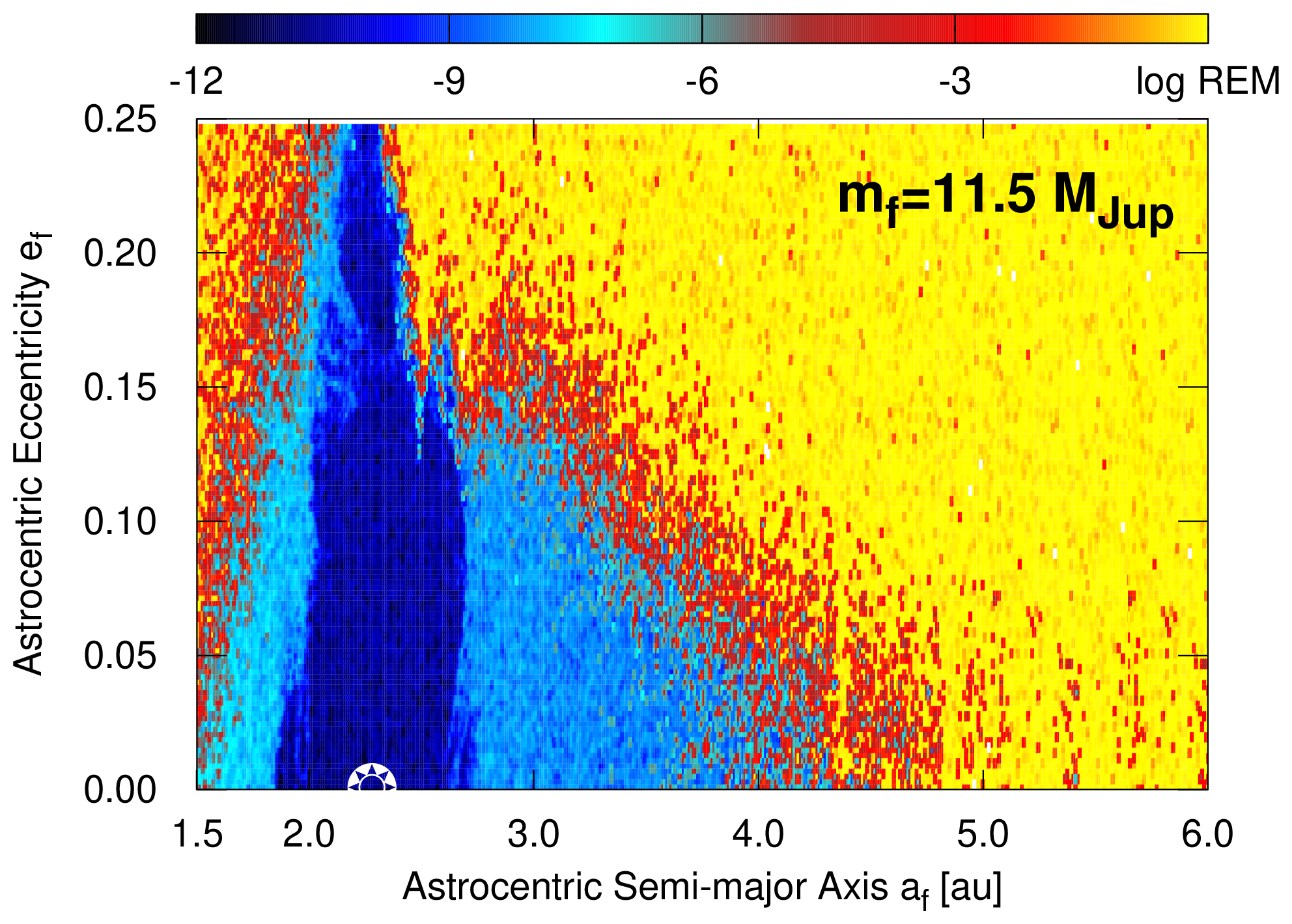}}
\hbox{\includegraphics[width=0.48\textwidth]{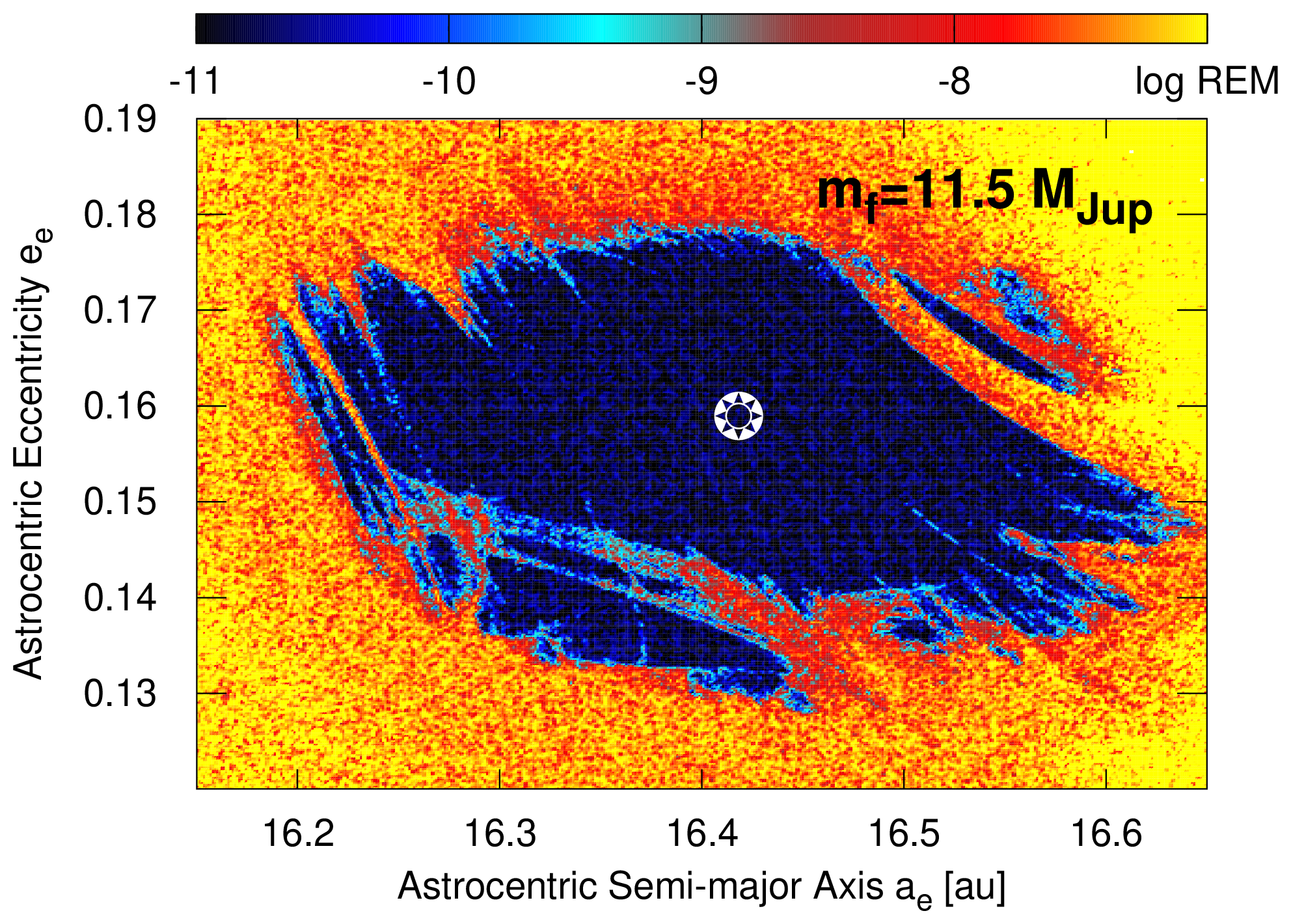}}
}
\centerline{
\hbox{\includegraphics[width=0.48\textwidth]{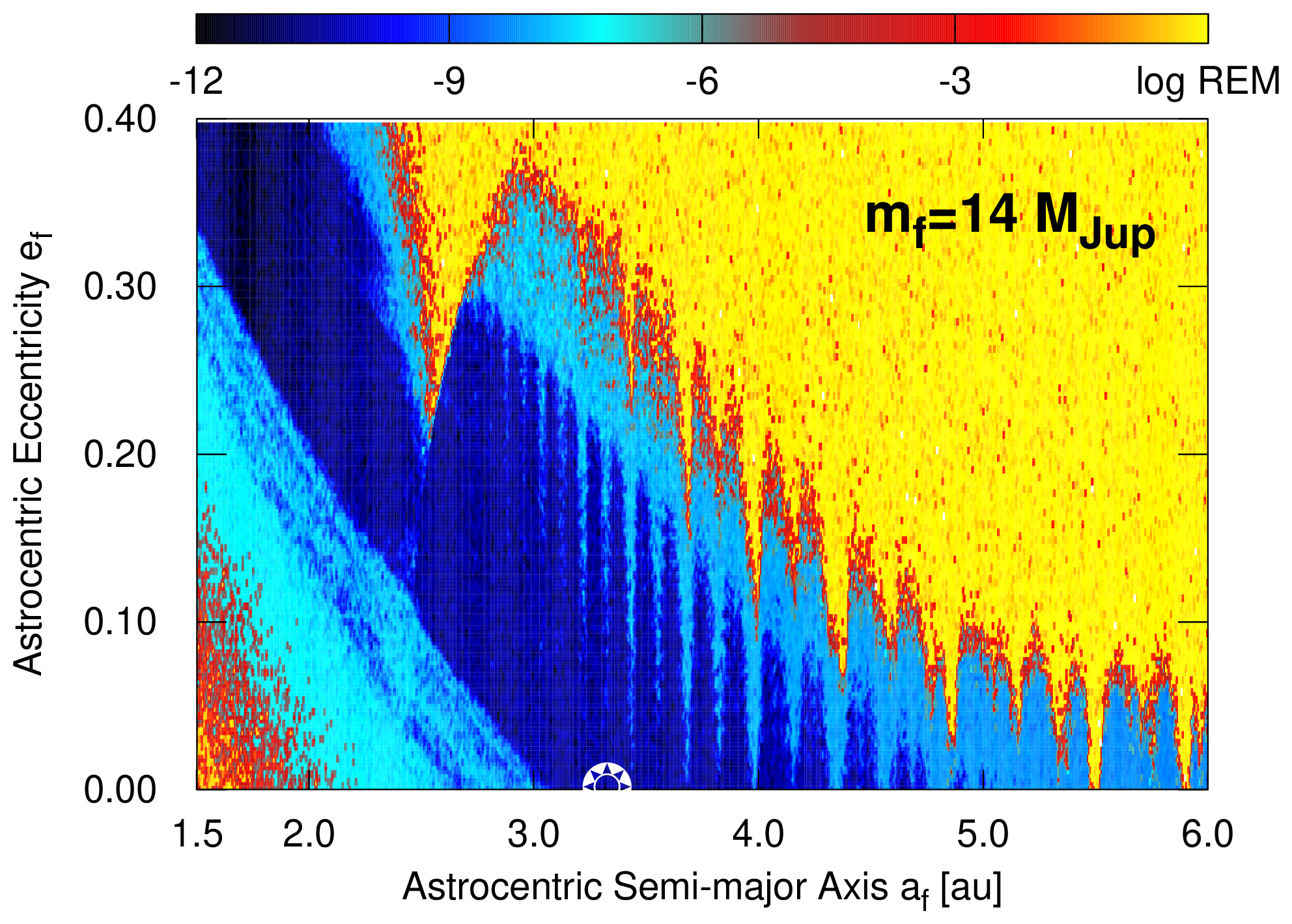}}
\hbox{\includegraphics[width=0.48\textwidth]{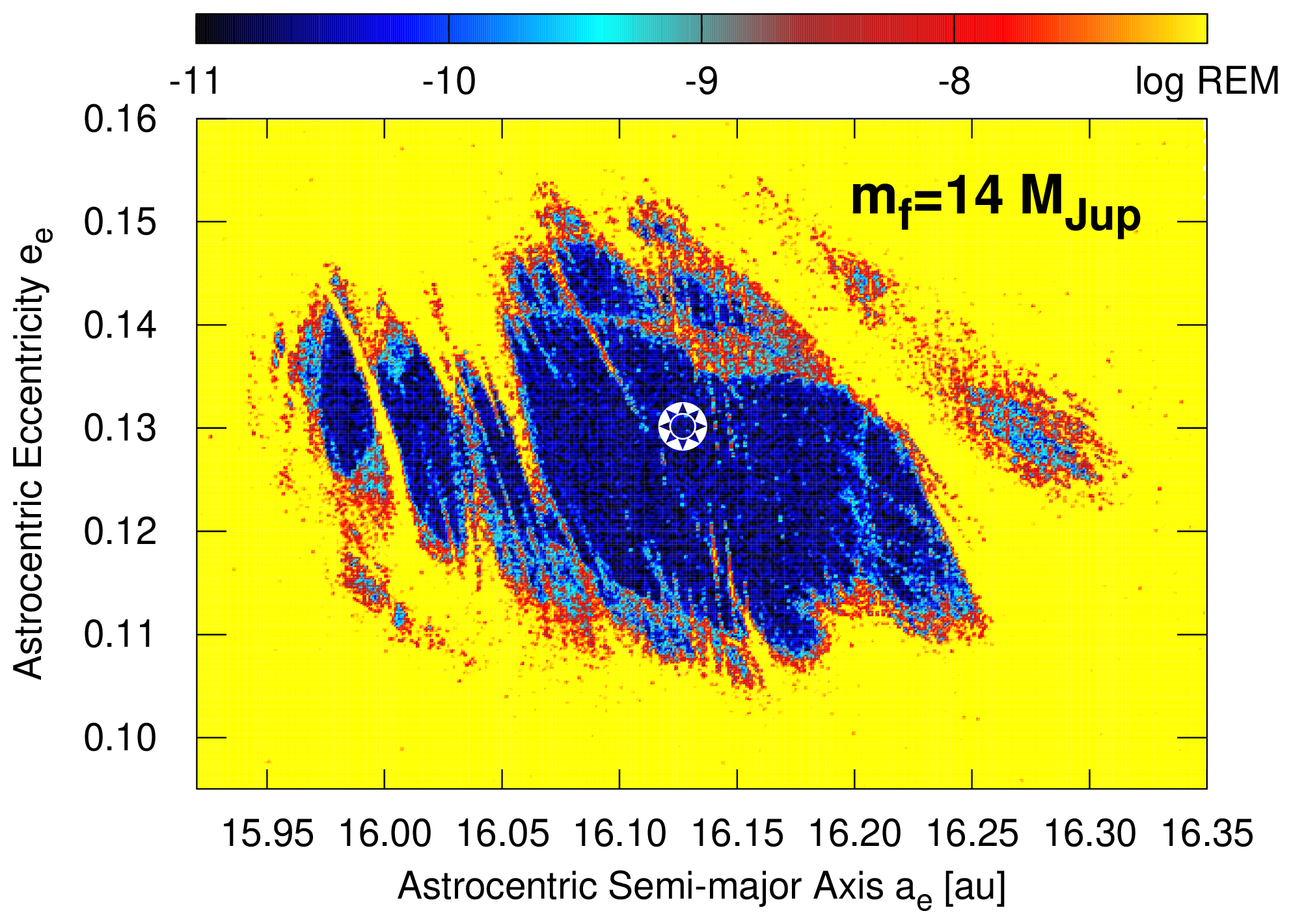}}
}
\centerline{
\hbox{\includegraphics[width=0.48\textwidth]{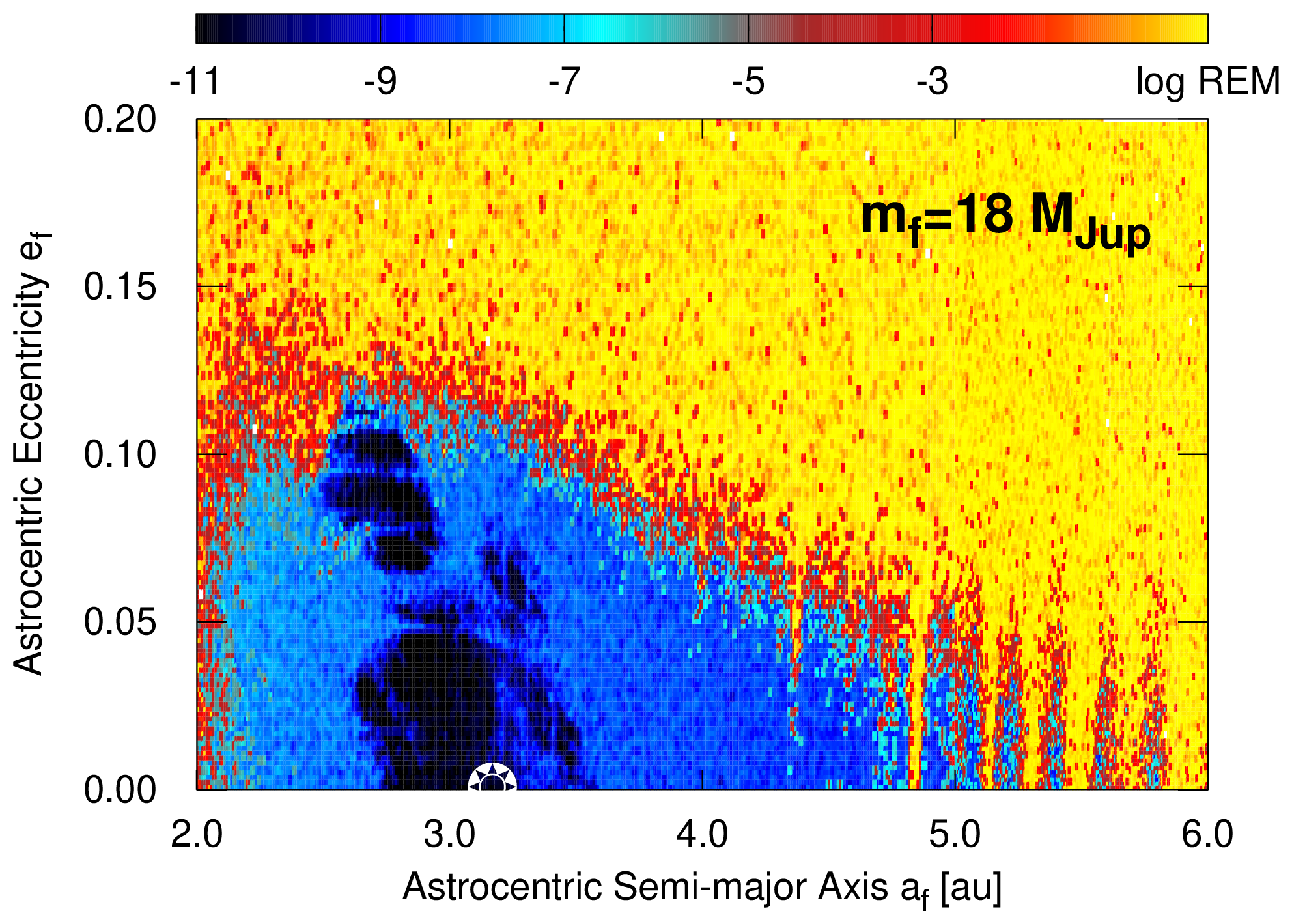}}
\hbox{\includegraphics[width=0.48\textwidth]{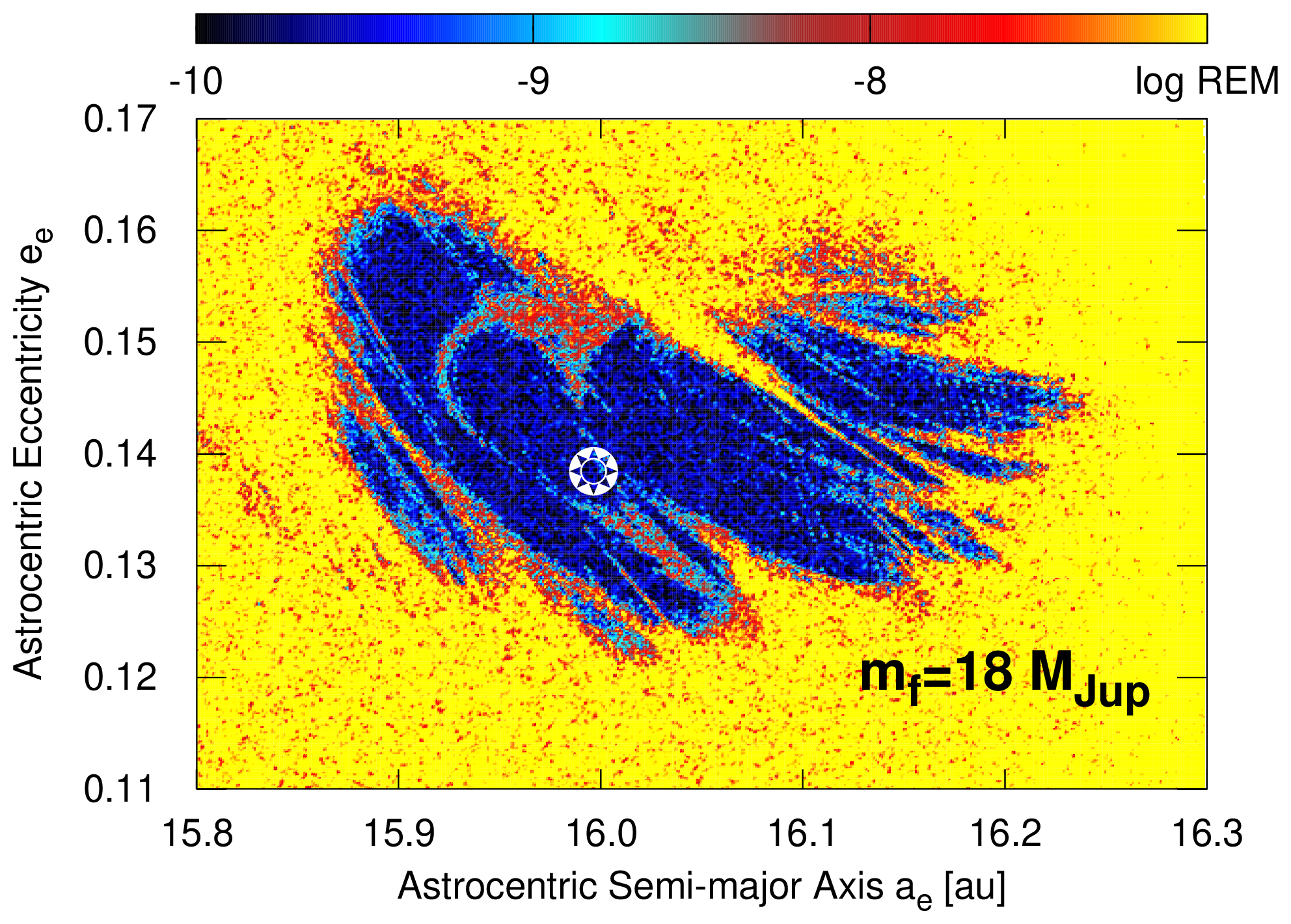}}
}
\caption{
Dynamical stability maps for the five-planet system. Left: scans in the $(a_{\rm f},e_{\rm f})$ plane for planet~f; right: corresponding scans in the $(a_{\rm e}, e_{\rm e})$ plane for planet~e, tracing the stability of the outer Laplace-type resonance chain. Each row corresponds to a representative solution with $m_{\rm f}=11, 11.5, 14,$ and 18\mJup{}  (Models~11, 115, 14, and 18 (\edtab{tab:EX-models}); the filled circles indicate the nominal configurations. Dark colours correspond to long-term stable evolution, whereas bright colours indicate strongly chaotic and unstable evolutions. 
}
\label{fig:dynmaps}
\end{figure*}
\begin{figure*}
\centerline{
\vbox{
\hbox{\includegraphics[width=0.8\textwidth]{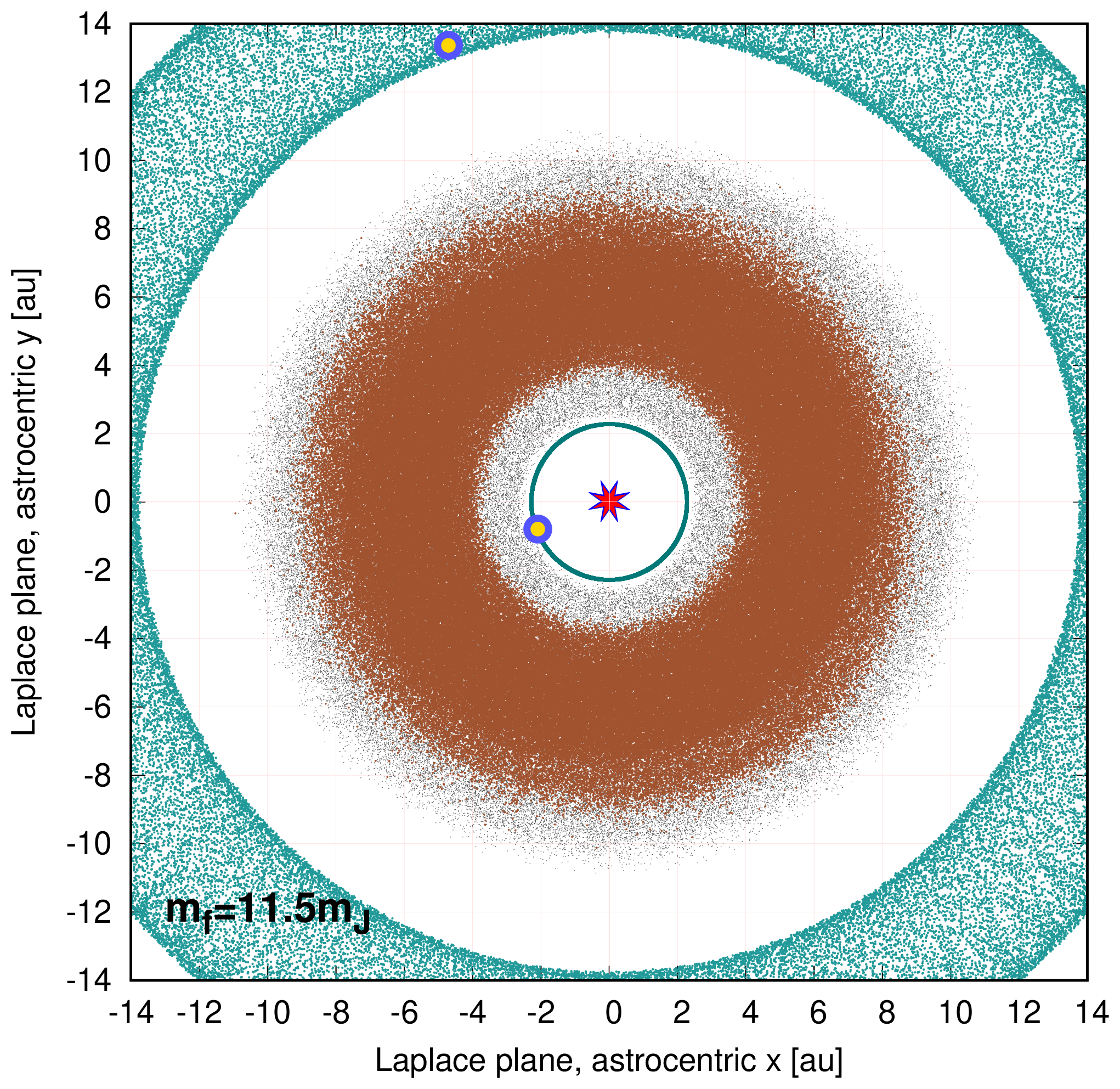}}
}
}
\caption{ 
Long-term $N$-body simulations of the inner warm debris disk. Planet~f has a mass of $11.46$\mJup (Model~115 in \edtab{tab:EX-models}) and an semi-major-axis of $2.27$\,au, close to the center of the dominant GaiaPMEX solution family. The initial disk (gray points, $\simeq 10^6$ particles) spans from $\sim$2.5-10\,au. The brown points show the $\simeq 300,000$ particles in the disk at $t_0=35$\,Myr. Large symbols mark the position of \host{} and planets~e and~f at the final epoch, while the green dots trace the locations of planets~e and~f sampled throughout the integration.
%
}
\label{fig:disk11}
\end{figure*}

%
\section*{\centerline{\Large Methods}}
\subsection*{\centerline{Absolute Astrometry}}
\subsubsection*{GaiaPMEX }
\label{sec:pmex}
We use two complementary absolute-astrometry diagnostics to search for unresolved companions: the Hipparcos-Gaia proper-motion anomaly (PMa; \citealp{Brandt2021,Kervella2022}) and the Gaia DR3 renormalized unit weight error (RUWE). 
We also use the astrometric excess noise (AEN) as a consistency check, but adopt RUWE as the primary Gaia scatter indicator. 

GaiaPMEX \citep{Kiefer2025a} forward-models the expected distributions of RUWE (and AEN) and the PMa under a null hypothesis that includes instrumental/processing astrometric noise and, when relevant, known companions. 
The noise is characterized by an along-scan uncertainty $\sigma_{\rm AL}$ and a calibration term $\sigma_{\rm calib}$. The correction strongly depends on the star’s G-magnitude and Bp-Rp color index \citep{Lindegren2021,Kiefer2025a}. 
For a given target star, GaiaPMEX draws Monte Carlo realizations of astrometric time series consistent with the star’s photometry and scanning-law statistics. Then it computes the corresponding RUWE, AEN, and PMa distributions under the null model, and evaluates tail probabilities for the observed values. $N\sigma$ significances correspond to one-sided (right-tail) probabilities converted to a Gaussian-equivalent significance. 

For \host{} (G=5.91, BP-RP=0.394), GaiaPMEX yields $\sigma_{\rm AL}=0.037$\,mas and $\sigma_{\rm calib}=0.246$\,mas. Under the single-star (noise-only) hypothesis, the predicted RUWE is $0.954 \pm 0.155$, while Gaia DR3 reports a RUWE of 1.47. This highlights the importance of the new estimate of the Gaia noises. The PMa generated from noise only would be $0.133 \pm 0.080$\mas/yr. The observed RUWE and PMa ($0.399 \pm 0.061$\,mas\,yr$^{-1}$; \citep{Kervella2022}) are inconsistent with the single-star expectation at $3.3\sigma$ (RUWE) and $2.7\sigma$ (PMa), indicating excess astrometric motion of physical origin.

GaiaPMEX then computes mass-semi-major-axis confidence maps for an inner companion by evaluating the likelihood of the observed RUWE and PMa given (i) the null model and (ii) a single companion on a Keplerian orbit. It marginalizes over nuisance orbital angles using the same priors as in GaiaPMEX \citep{Kiefer2025a}. The resulting RUWE-only, PMa-only, and joint RUWE+PMa maps are shown in Fig.~\ref{fig:fig1}.

\subsubsection*{Checking for source confusion and stellar variability}

To exclude spurious RUWE inflation from source confusion, we verified the Gaia DR3 image-parameter determination diagnostics: \texttt{ipd\_frac\_multi\_peak} and \texttt{ipd\_gof\_harmonic\_amplitude}. These values, respectively 0 and 0.044, indicate  no evidence for multiple PSF peaks and minimal systematic PSF-shape variability \citep{Holl2023}. 

We also considered whether star-related phenomena could produce the required photocentre motion. First, the amplitude of photocentre shifts due to stellar jitter caused by spots, plages, granulation, non-radial oscillations, is expected to be at most a few  tens of $\mu$AU \citep{Eriksson2007}, corresponding to $<$0.001\,mas at HR8799's distance. Second, magnetic activity is not expected for an early-type star such as HR 8799. Moreover, any point-like source close to the stellar surface capable of producing a 0.25-mas astrometric signal would need to outshine the star while orbiting at $\simeq 1.5 R_{\rm star}$. We are not aware of any plausible scenarios involving faculae, flares, or other surface features in early-type stars that could satisfy these conditions. 
Order-of-magnitude estimates further show that producing a non-negligible photocentre excursion relative to 0.25 mas would require a surface-brightness asymmetry comparable to (or exceeding) the stellar flux distributed over spatial scales of order the stellar radius. This is not expected for \host{}, a $\gamma$-Doradus pulsator exhibiting individual g-mode amplitudes $<$15\,mmag and peak-to-peak photometric variations $<$100\,mmag \citep{Sodor2014}, corresponding to fractional flux variations $<$10\%.

\subsubsection*{Accounting for the known planets \host{}bcde}
Because \host{} hosts four massive, wide-orbit planets, we include their contribution to the host star’s reflex motion in the null model.  We adopt the bcde orbital solution from \citep{Zurlo2022} (Model~1 of their Table~7) and propagate the system forward with an $N$-body integration from epoch 1998.83 to an epoch near Gaia DR3 reference epoch (2016.0) (\edtab{tab:EX-Model1}). We then forward-model the expected PMa and RUWE distributions, including the combined Keplerian reflex motion of bcde and the DR3 noise. In this four-planet null model, the PMa becomes consistent (within $>0.8-\sigma$) with the predicted distribution, whereas RUWE (and AEN) remain significantly ($>2.8-\sigma$) elevated (\edfig{fig:EX_Fig1}), showing that the PMa is dominated by the known planets (primarily \host{}e) while the RUWE/AEN excess requires an additional inner companion. We therefore subtract the predicted bcde contribution from the astrometric observables and recompute the companion mass-semi-major-axis constraints. The resulting confidence maps still favor an additional, close-in substellar companion (Fig.~\ref{fig:fig3}). 

\subsubsection*{Checking the impact of mutual Newtonian interactions on RUWE and PMa}
To assess whether neglecting mutual interactions among bcde biases the predicted RUWE or PMa, we ran two GaiaPMEX forward models: one using the osculating elements at epoch 1998.83, and one using the propagated osculating elements at epoch 2016.0. The resulting changes in the predicted RUWE and PMa distributions are 0.008 and ~0.003 mas/yr, respectively, much smaller than the uncertainty on the RUWE and PMA due to the noise (0.16 and 0.12\masyr, respectively). The mutual-interaction correction is therefore negligible when interpreting RUWE and PMa. We adopted the parameters of the osculating orbit at 2016.0 to describe the orbital motion of planets bcde in the present analysis. 

\subsubsection*{Combined Bayesian model}

Orbital elements of planet f were fit using MCMC sampler emcee \citep{Foreman-Mackey2013}, and upgraded to account for both Gaia, PMa, direct imaging and RV data (as in \citep{Lagrange2025}). We used in addition a model in which the orbital elements of planets bcde were fixed to the values reported in Table~\edtab{tab:EX-Model1}. The MCMC sampling used 100 chains run for 200,000 steps, with the first 25,000 steps discarded as burn-in.

The likelihood associated with absolute astrometry (PMa and RUWE) was computed with the GaiaPMEX code. For each sampled parameter set, a bootstrap procedure generates distributions of RUWE and PMa from the random dispersion of simulated astrometric time series. Their mean and standard deviation are then used to define the likelihood of the observations as:
\begin{align}
\ln {\cal L}_{\mathrm{PMEX}}
&= -\frac{1}{2}\Bigg[
\left(
\frac{ f(\mathrm{ruwe}_{\mathrm{obs}}) - \overline{f(\mathrm{ruwe}_{\mathrm{model}})} }
{ \sigma\!\left(f(\mathrm{ruwe}_{\mathrm{model}})\right) }
\right)^{2}
\;+\;
\sum_{i\in\{\mathrm{Ra},\mathrm{Dec}\}}
\frac{\left(\mathrm{PMa}_{i,\mathrm{obs}}-\overline{\mathrm{PMa}_{i,\mathrm{model}}}\right)^{2}}
{\sigma\!\left(\mathrm{PMa}_{i,\mathrm{model}}\right)^{2}+\epsilon\!\left(\mathrm{PMa}_{i,\mathrm{obs}}\right)^{2}}
\Bigg] \nonumber\\[4pt]
&\quad
-\frac{1}{2}\Bigg[
\sum_{i\in\{\mathrm{Ra},\mathrm{Dec}\}}
\ln\!\left(\sigma\!\left(\mathrm{PMa}_{i,\mathrm{model}}\right)^{2}
+\epsilon\!\left(\mathrm{PMa}_{i,\mathrm{obs}}\right)^{2}\right)
\;+\;
\ln\!\left(\sigma\!\left(f(\mathrm{ruwe}_{\mathrm{model}})\right)^{2}\right)
\Bigg].
\end{align}

Here, $f$ is taken from \citep{Kiefer2025a} and converts RUWE to the one-third power of the unbiased estimator of variance a posteriori (UEVA) of the Gaia residuals.

A one-dimensional contrast curve derived by \citep{Dallant2023} from multi-epoch SPHERE direct-imaging data was converted into a one-dimensional mass–separation curve using the evolutionary models of \citep{Baraffe2015}. At each step we compared the sampled companion mass to the detection limit at the corresponding separation; steps lying above the imaging detection limit were rejected.

The RV data were also incorporated as upper limits, since no planetary signal is detected. High amplitude ($\geq $ 1km/s) RV variations are present, and are attributed to strong stellar pulsations, consistent with previous reports \citep{Lagrange2013,Grandjean2020}. The standard deviation of the series, $\sigma_{\mathrm{obs}}$, was computed. To test the detectability of a given model companion, we used either a rms-based detectability criterium or the Local Power Analysis (LPA) approach. Both approaches are detailed and compared in  \citep{Meunier2012} and used in e.g. \citep{Lagrange2013}. For each walker, and at each step (hence a given mass and orbital elements), we simulated an RV time series sampled at the same epochs as the observations and computed  its rms.
 In the rms-based approach, the step is rejected if the model rms is greater than  $ 1\times \sigma_{obs}$ (\citep{Meunier2012} and \citep{Lagrange2013}). 
In the LPA approach, a periodogram-based detection-limit curve was computed, that better accounts for the high-frequency pulsations than the rms approach; a step was rejected when the sampled $(a, m)$ lay above the LPA limit curve. The LPA and rms-based approaches yield consistent posteriors, with the highest posterior density ($ 75-83 \%$) at $\sim$2-3\,au and with masses $\sim$10-14\,\mJup, a low density solution at (0.8au, 40\,\mJup) and a secondary mode contributing about $14-20 \%$ near (0.5au, 80\,\mJup). The figures and values provided in the paper correspond to the results obtained with the rms mode.

\subsection*{\centerline {Stable five-planet model with outer Laplace resonance chain}}

\subsubsection*{Adiabatic MMR tuning}
Given that prior analysis of the available relative astrometry data did not detect the inner companion, we assume that the reported outer resonant or near-resonant Laplace 8e:4d:2c:1b chain remains intact in the presence of this inner companion. The narrow, stable equilibrium-like resonant architecture of the \host{}bcde system is a primary factor in the long-term stability \citep[e.g.,][]{Zurlo2022,Wang2018}.  Therefore, we consider the four-planet Model~1 of  \citep{Zurlo2022} 
as our baseline outer architecture. We assume that all orbits are in the Laplace plane and fix ${\rm M}_{\star}=1.47\,{\rm M}_{\odot}$ \citep{Sepulveda2022}.

Inserting a massive planet into the bcde configuration generally shifts the system’s mass center and causes uncontrolled changes to osculating orbital elements that define the Laplace resonance in the parameter space, leading to an apparent instability. To construct a self-consistent five-planet initial condition that accurately encodes the \host{}bcde resonance chain in the configuration with an additional planet, we employ a heuristic algorithm based on the planetary migration paradigm \citep{Cresswell2008}, previously successfully used to  model the relative astrometry of the \host{}bcde system \citep[MCOA,][]{Gozdziewski2014,Gozdziewski2018}. We expand the orbits and integrate the $N$-body equations of motion while applying weak dissipative forces. This leads to adiabatic damping of semi-major axes and eccentricities, enabling convergent migration and resonance re-capture, which stabilizes the system orbits.
 
To initiate  migration, the five-planet system composed of planet~f and the outer resonant subsystem was uniformly rescaled by increasing all semi-major axes by a common factor (typically $\simeq 20$), with randomized orbital phases and small initial eccentricities, in order to suppress short-timescale, close encounters during the adiabatic evolution. We explored masses of 10 to 18\mJup and a reference semi-major axis of planet~f, $a_{\rm f}$ near $2.5$\,au, within the $1\sigma$ GaiaPMEX dominant solutions. We also considered companions with 40, 60 and 80\mJup and $a_{\rm f}<1$\au, corresponding to the secondary-modes solutions.

After the orbits expansion, parametrized slow, inward migration and weak eccentricity damping were applied to each bcde planet, individually, according to 
\begin{equation}
a(t) = a(t_0) \exp(-t/\tau_{{\rm a}}),
\quad
e(t) = e(t_0) \exp(-t/\tau_{{\rm e}}),
\end{equation}
 where $a(t_0)$ and $e(t_0)$ denote the initial values and $\tau_a$ and $\tau_e$ are the migration and damping timescales. $\tau_{e}$ and $\tau_{\rm a}$ were chosen so that the ratio $\tau_{\rm a}/\tau_{\rm e} \simeq 2$. They serve as control parameters to enforce near-conservative evolution, rather than to model any specific physical migration mechanism. The timescales were chosen to be much longer than the MMR libration periods $\tau_{\rm lib} \propto 10^{2} P_{\rm orb}$, where $P_{\rm orb}$ is the orbital time scale, in the two-body MMRs of giant planets ($\tau_{a} \gg 10^5 \tau_{\rm lib}$), ensuring adiabatic evolution and suppressing impulsive perturbations that would otherwise disrupt resonant locking \citep{Batygin2015}.

\subsubsection*{Parametric search and resonance identification}

For each trial $(m_{\rm f}, a_{\rm f})$, initially expanded synthetic systems were integrated forward under dissipation for up to  $\tau_{\rm max}$, covering several $e$-folding times to allow mean-motion resonance capture and settling of the resonant angles. We found the capture of \host{}bcde into the Laplace chain robust, as it occurs with a high probability (up to several tens of percent of initial configurations).

Osculating orbital elements at epoch $t_0$ corresponding to the observed $a_{\rm e} \simeq 16$\,au (Model~1 in \citep{Zurlo2022}) were then selected. These models were subsequently validated through direct long-term $N$-body integrations without dissipative forces and dynamical maps with fast indicators.

The Laplace resonance capture is assessed based on (i) near-commensurate 2:1 period ratios for adjacent pairs in the outer chain, (ii) sustained libration of all relevant two-body resonant critical angles within a few tens of degrees,  (iii) libration of the associated four-body Laplace-type critical angle with the semi-amplitude less than 45~degrees, as important for the long-term stability, and (iv) equilibrium eccentricities close to respective values in Model~1 \citep{Zurlo2022}. See Fig.~\ref{fig:orb_evol11}, \edfig{fig:EX-els1444} and  \si{} for details.

\subsubsection*{Stability analysis and numerical integration}

Long-term orbital evolution of the resonant five-planet solutions obtained with MCOA was assessed through $N$-body integrations without dissipation, covering timescales exceeding the system age. We employed the symplectic integrator SABA4 \citep{Laskar2001} and an independent, higher-order symplectic scheme as a cross-check \citep{Beust2024}. Stability of the system was quantified by the absence of close encounters, bounded evolution of semi-major axes and eccentricities, sustained low-amplitude libration of all relevant critical angles, and a maximal Lyapunov exponent near zero, consistent with quasi-periodic evolution (\si).

To quantify stability robustness of the model in the phase space, we computed dynamical maps around representative solutions within the dominant posterior mode (\edtab{tab:EX-models}) by scanning the $(a,e)$ plane of selected planets and quantifying system stability using the Reversibility Error Method (REM), a fast Lyapunov exponent-like indicator \citep{Panichi2017}. For each grid point, $\log{\rm REM}$ was used to distinguish regular (stable) from chaotic (unstable) evolution (Fig. \ref{fig:dynmaps}). Details of the dynamical map construction
are provided in \si{}.

\bibliography{ms}
\section*{Acknowledgments}
 This project has received funding from the European Research Council (ERC) under the European Union's Horizon 2020 research and innovation program (COBREX; grant agreement \#885593). K.G. acknowledges the support and CPU resources provided by the Pozna\'n Supercomputing and Networking Centre, Poland (PCSS, project PL0406-01). A.Z. acknowledges support from ANID -- Millennium Science Initiative Program -- Center Code NCN2024\_001 and Fondecyt Regular grant number 1250249.

\noindent {\bf \large Data availability}

\noindent All observational data in this work are available through public data archives, or published, or provided in Additional Data. 

\noindent {\bf \large Code availability}

\noindent Detailed descriptions of the codes used in the present paper can be found in the provided references.

\noindent {\bf \large Author information}
A-M Lagrange led the project, the GaiaPMEX analysis, the interpretation of the various observational data, the disk analysis, and the writing of the paper. 
K. Go\'zdziewski led and conducted all the dynamical studies and wrote the associated text.
F. Kiefer upgraded the GaiaPMEX code to allow for subtracting the contribution of the outer planets. He contributed to the writing of the Methods on Gaia use, together with AML. 
H. Beust and P. Thebault contributed to the dynamical analysis.
P. Rubini performed the MCMC analysis.
A. Zurlo contributed to the general discussion on the system. A. Boccaletti contributed to the analysis of the inner disk interpretation.
All authors commented on the manuscript.

\noindent {\bf \large Ethics declarations} \noindent The authors declare no competing interests.

 \clearpage

 \clearpage
\newpage

\setcounter{figure}{0}
\setcounter{table}{0}
\renewcommand{\thefigure}{\arabic{figure}}
\renewcommand{\thetable}{\arabic{table}}

\renewcommand{\figurename}{Extended Data Figure}
\renewcommand{\tablename}{Extended Data Table}

\section*{Extended Data}


\subsection*{Extended Data Figures}
%

\begin{figure*}
\centerline{
\hbox{
\hbox{\includegraphics[width=1.\textwidth]{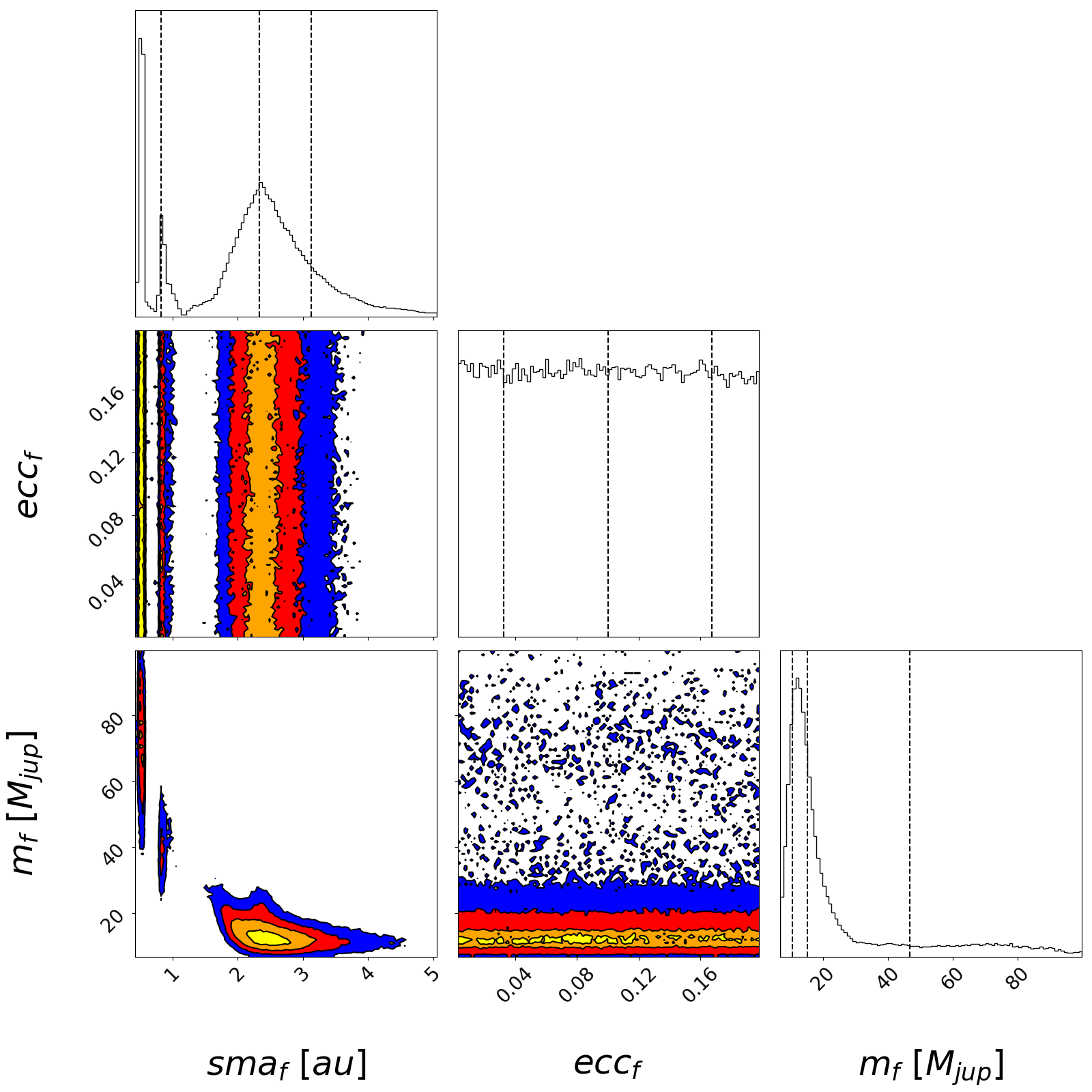}}
}
}
\caption{
{\bf Corner plot of the MCMC fit of the Gaia and Gaia-Hipparcos PMa data, taking into account the RV and HCI detection limits.} The colors represent increasing likelihood, blue: $2\sigma$, red: $1.5\sigma$, orange: $1\sigma$, yellow: $0.5\sigma$. The priors are provided in \edtab{tab:EX-MCMCpriors}.
}
\label{fig:EX_CORNER}
\end{figure*}
\clearpage

 \clearpage

\begin{figure*}
\centerline{
\vbox{
\hbox{\includegraphics[width=0.8\textwidth]{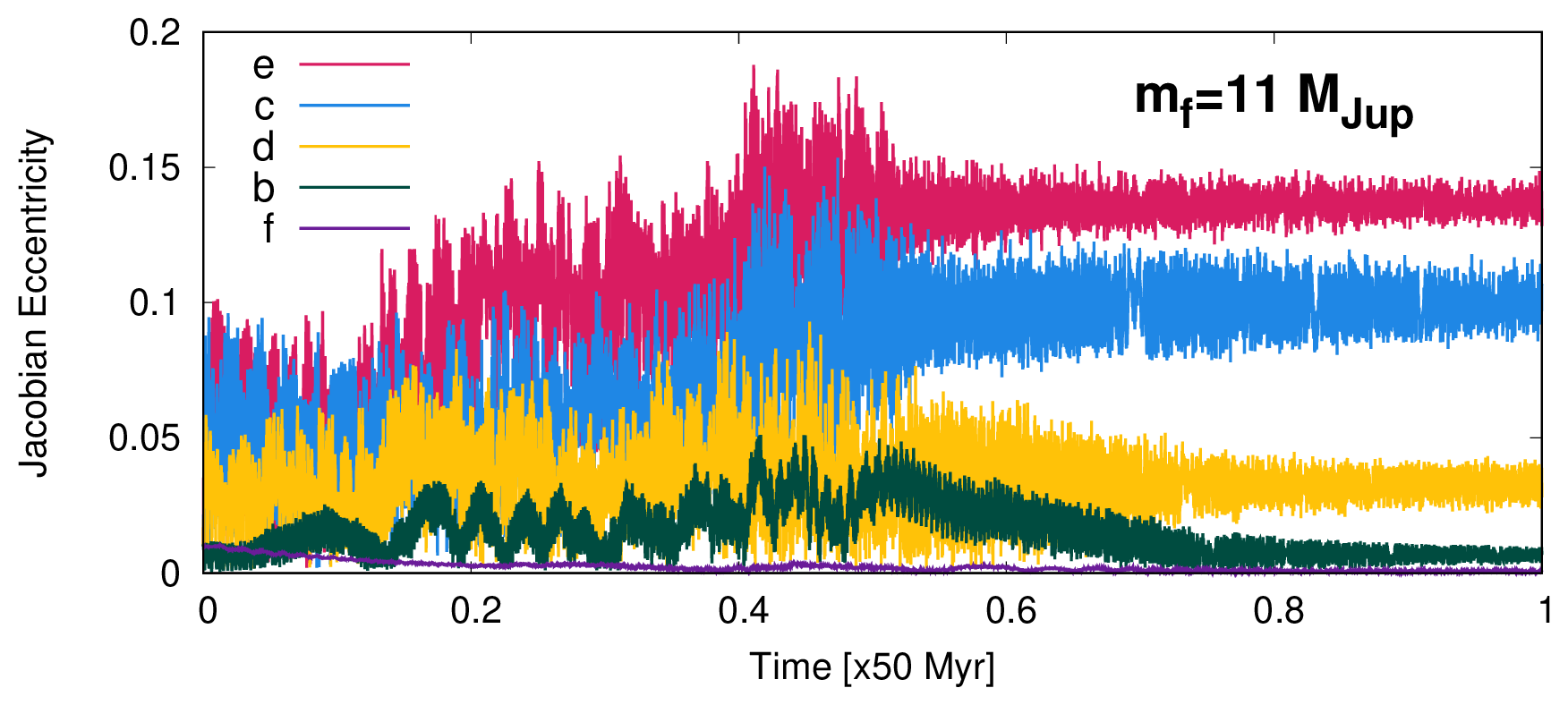}}
\hbox{\includegraphics[width=0.8\textwidth]{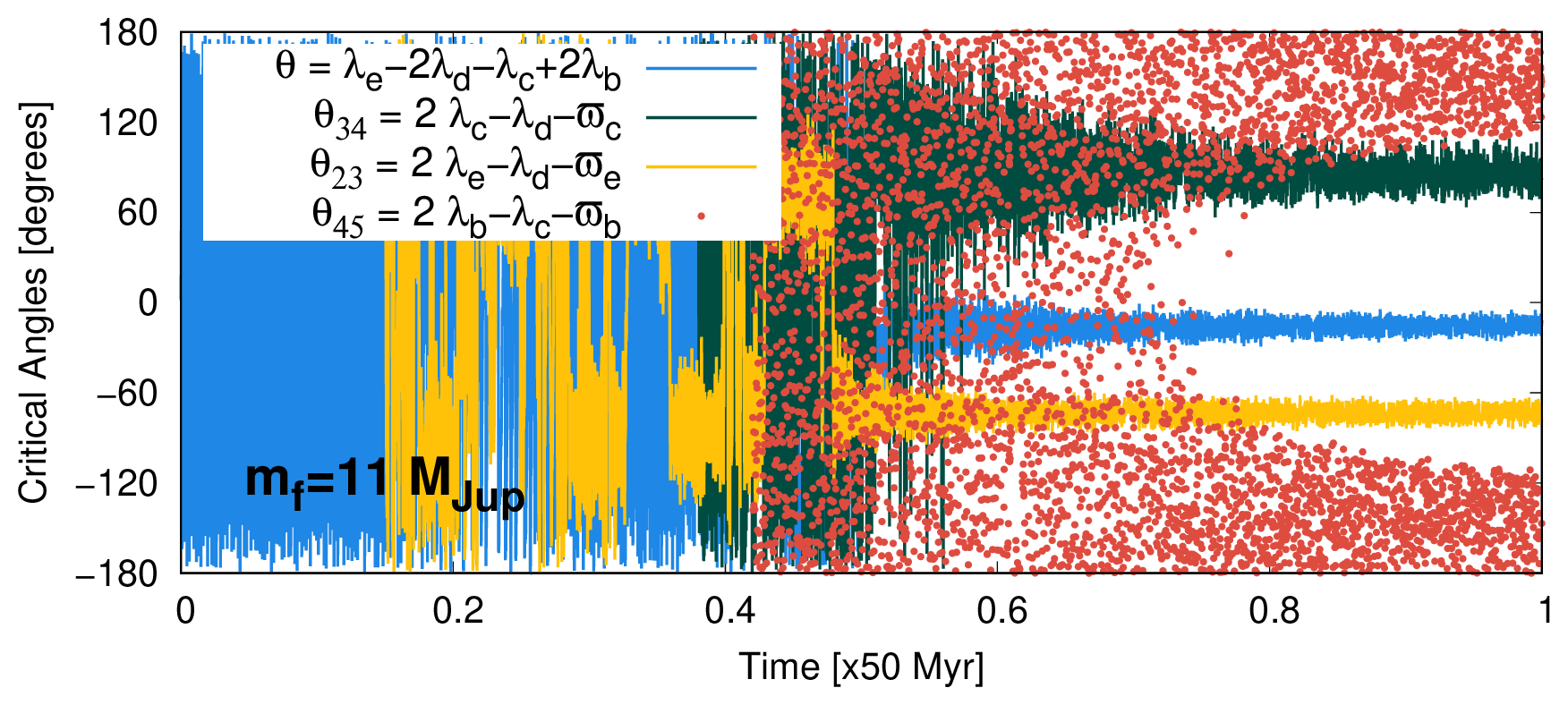}}
}
}
\caption{ 
{\bf Adiabatic, generic migration of the five-planet \host{} system with a  11\mJup companion initiated at $\simeq$ 3\,au}. 
{\em Top:} Eccentricity evolution over 50\,Myr. Initial eccentricities are $e_i = 0.01$ for the five planets. After $\simeq 25$~Myr, the eccentricities converge to equilibrium values characteristic of the Laplace MMR chain. 
{\em Bottom}: Temporal evolution of critical angles of the Laplace argument $\theta$ and three of six 2-body critical angles in the outer bcde system. After  $\simeq 25$-$30$~Myr, all those critical angles librate, indicating the MMR capture. In these settings, bcde systems similar to Model~1 with $a_{\rm e}\simeq 16$\au in \citep{Zurlo2022} appear around the final $\simeq 90$\% of the integration interval.  (Method and SI).
}
\label{fig:EX-migration}
\end{figure*}

 \clearpage

\begin{figure*}[h]
\centerline{
\hbox{
\hbox{\includegraphics[width=0.33\textwidth]{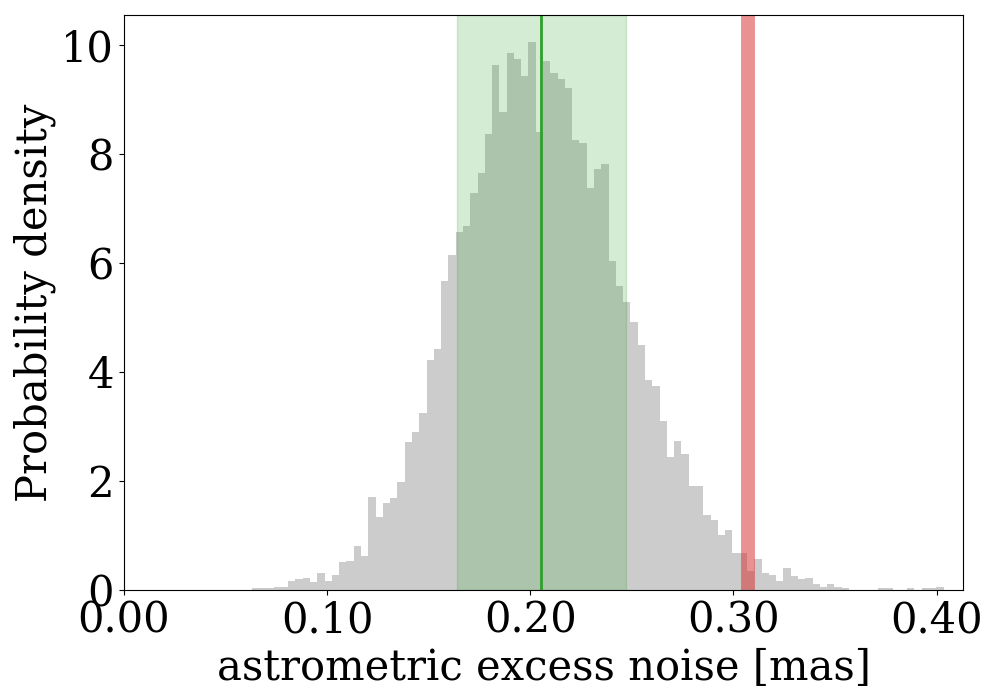}}
\hbox{\includegraphics[width=0.33\textwidth]{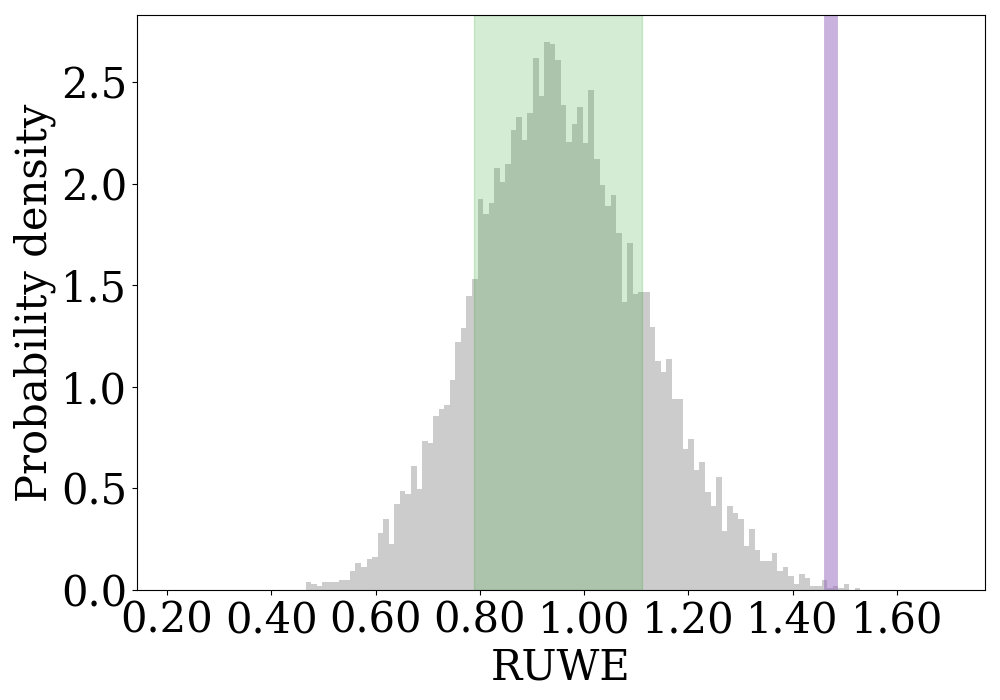}}
\hbox{\includegraphics[width=0.33\textwidth]{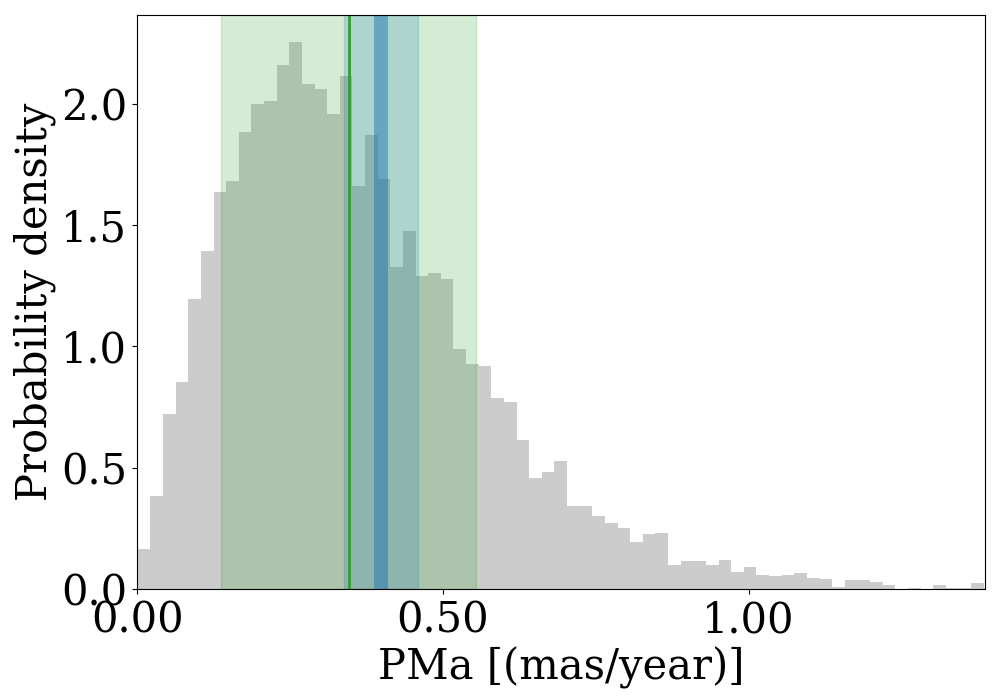}}
}
}
\caption{
{\bf Expected distributions of Gaia/Hipparcos astrometric indicators under the four-planet (bcde) null model}. Distributions of (left) AEN, (middle) RUWE, and (right) Hipparcos--Gaia PMa predicted by GaiaPMEX when including the reflex motion induced by the four known planets bcde (see text) and DR3 noise appropriate for the star’s $G$ and $BP-RP$. Vertical lines indicate the observed values. The PMa is consistent with the bcde model, whereas RUWE remains significantly elevated, indicative of the presence of an additional companion.
 }
\label{fig:EX_Fig1}
\end{figure*}
 \clearpage

\begin{figure*}
\vbox{
\centerline{
\hbox{\includegraphics[width=0.5\textwidth]{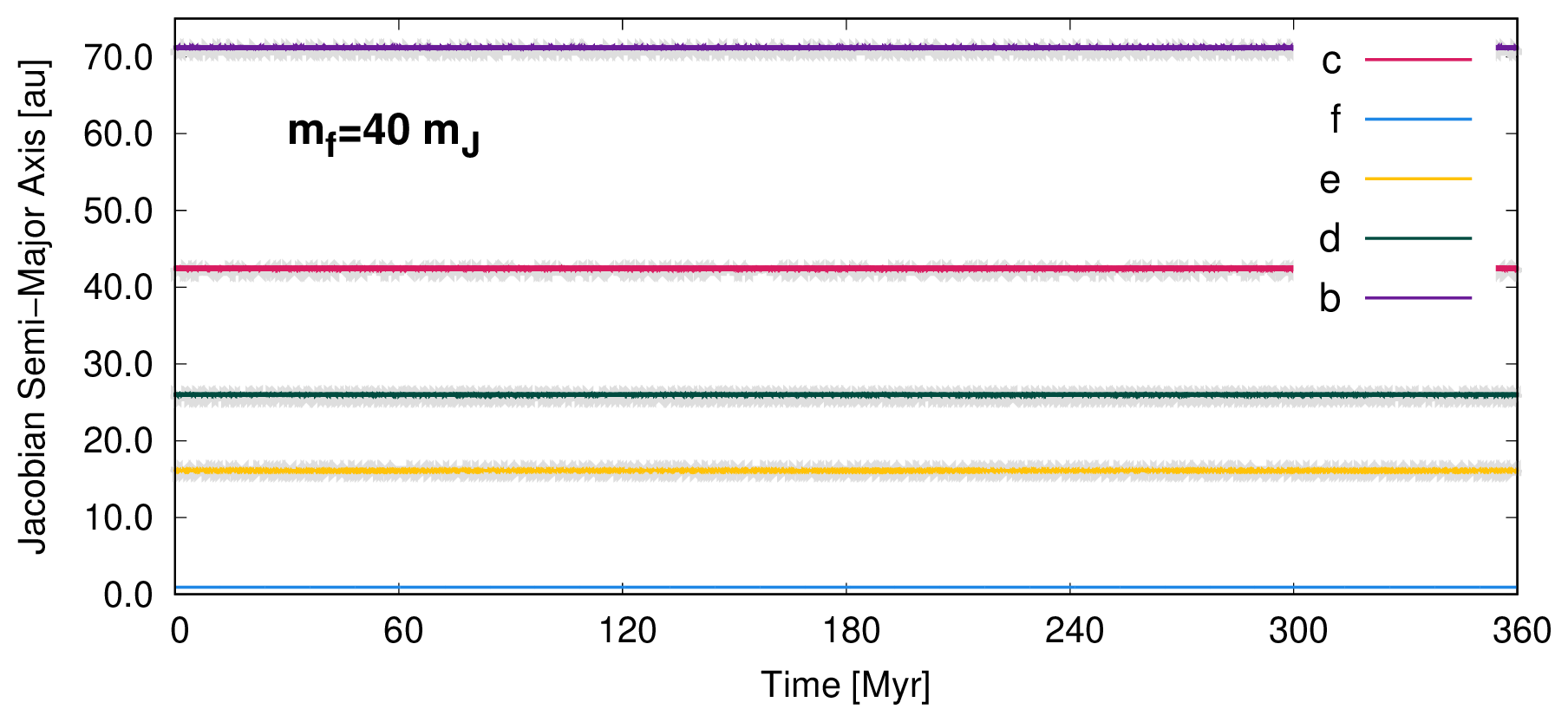}}
\hbox{\includegraphics[width=0.5\textwidth]{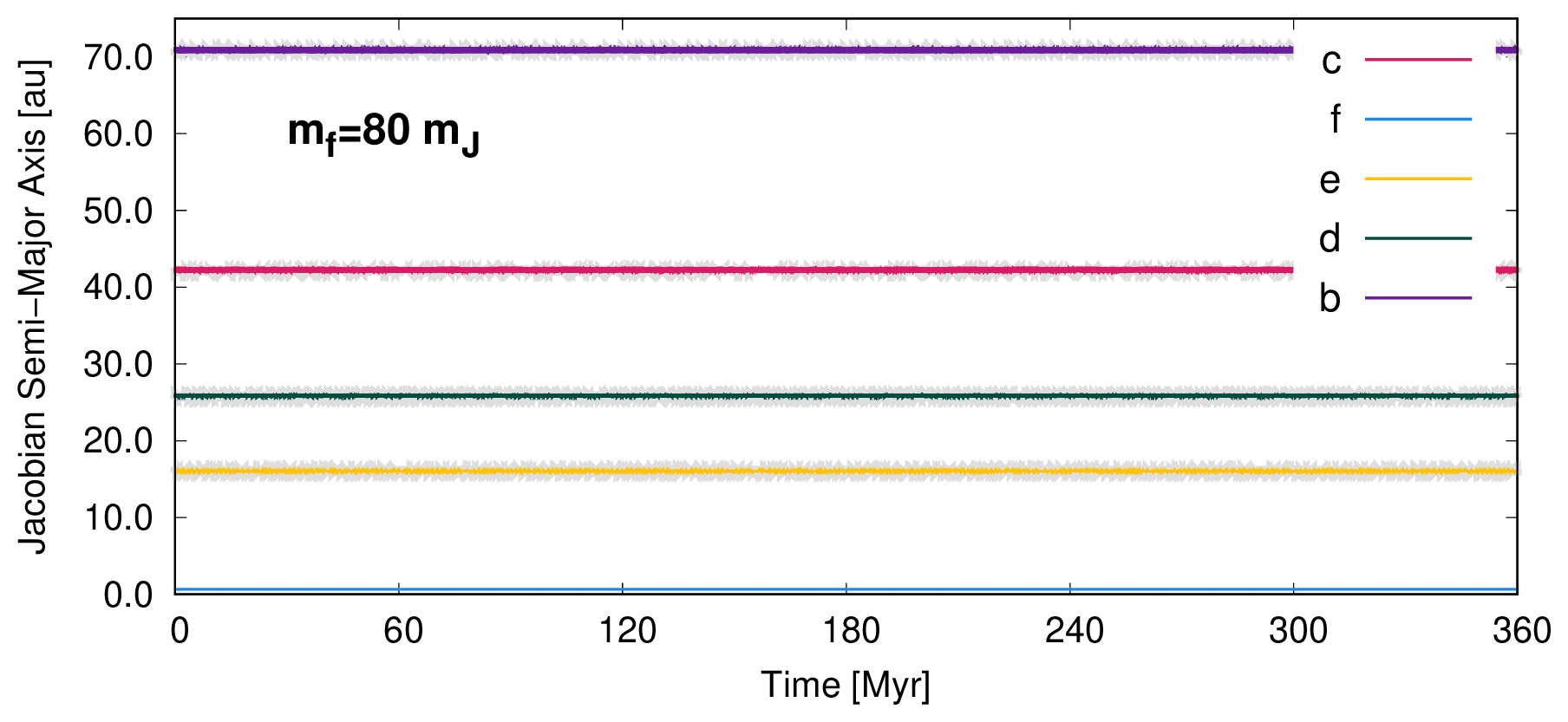}}
}
\centerline{
\hbox{\includegraphics[width=0.5\textwidth]{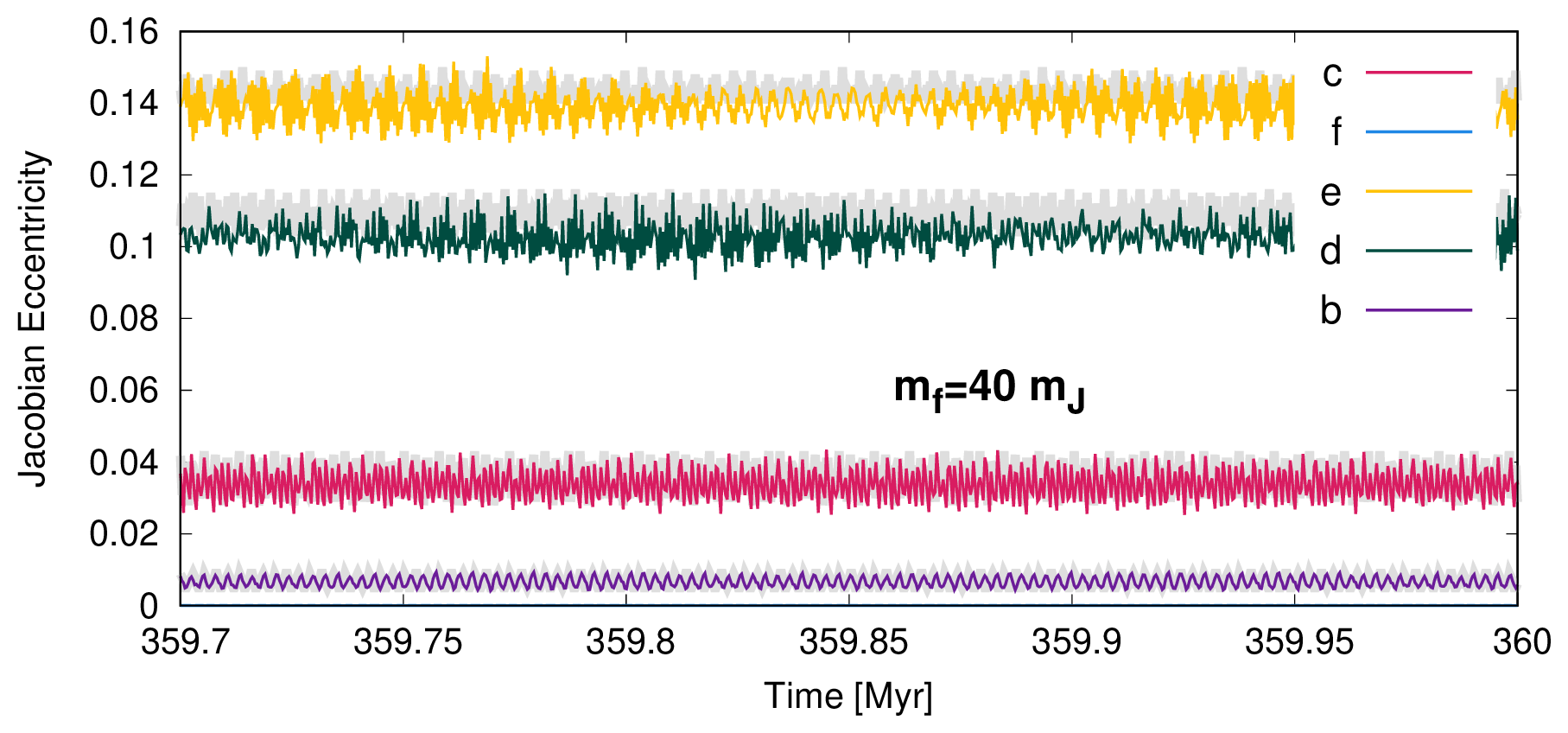}}
\hbox{\includegraphics[width=0.5\textwidth]{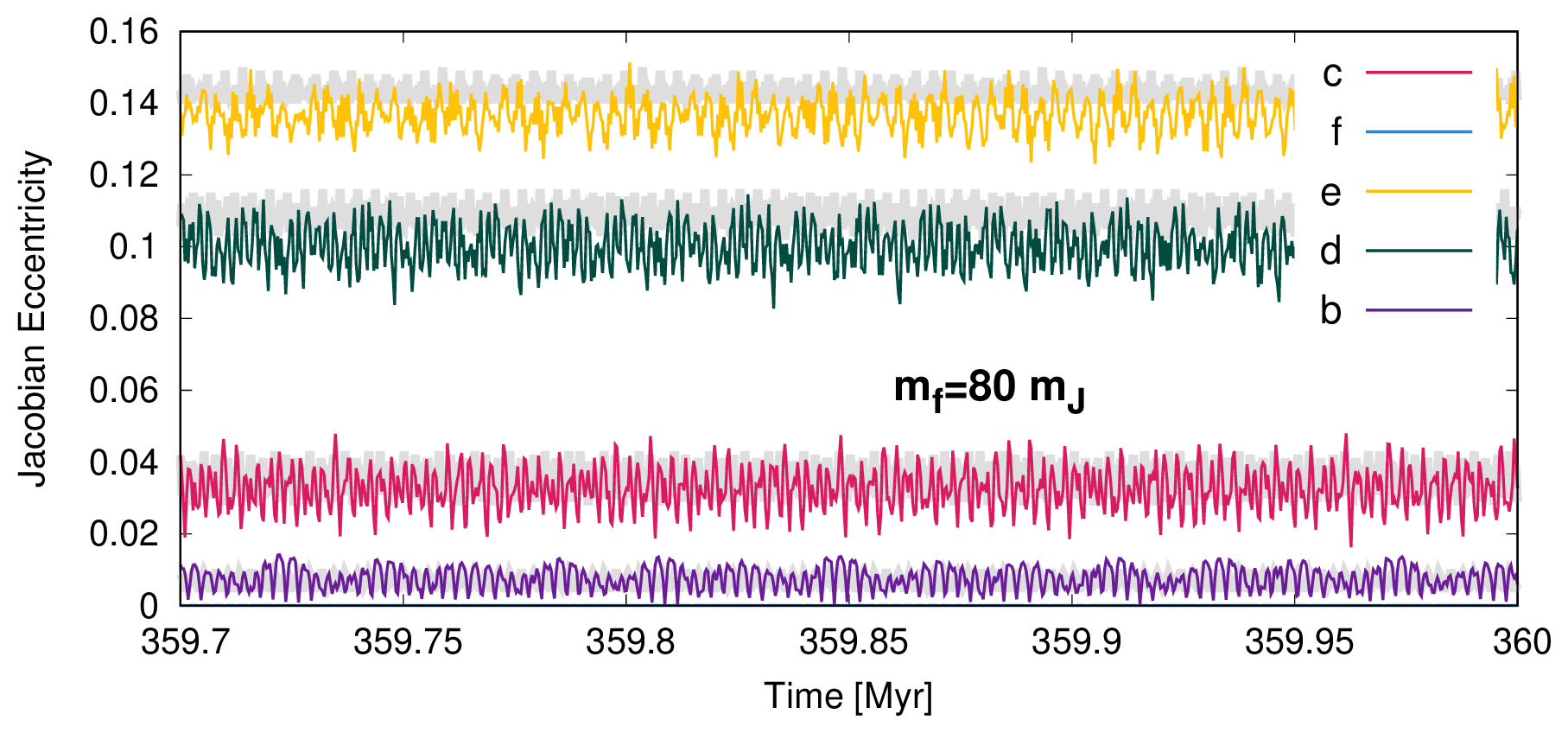}}
}
\centerline{
\hbox{\includegraphics[width=0.5\textwidth]{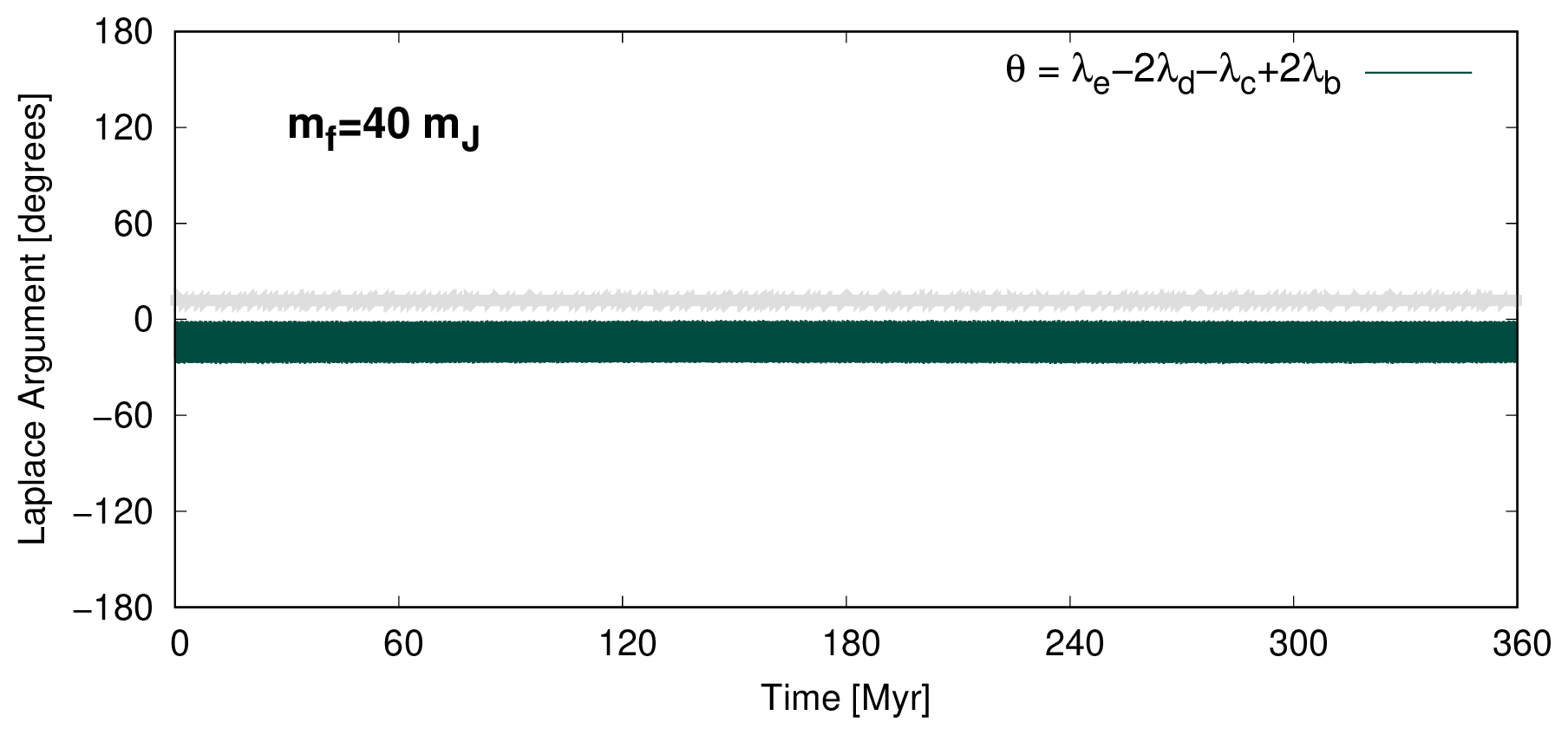}}
\hbox{\includegraphics[width=0.5\textwidth]{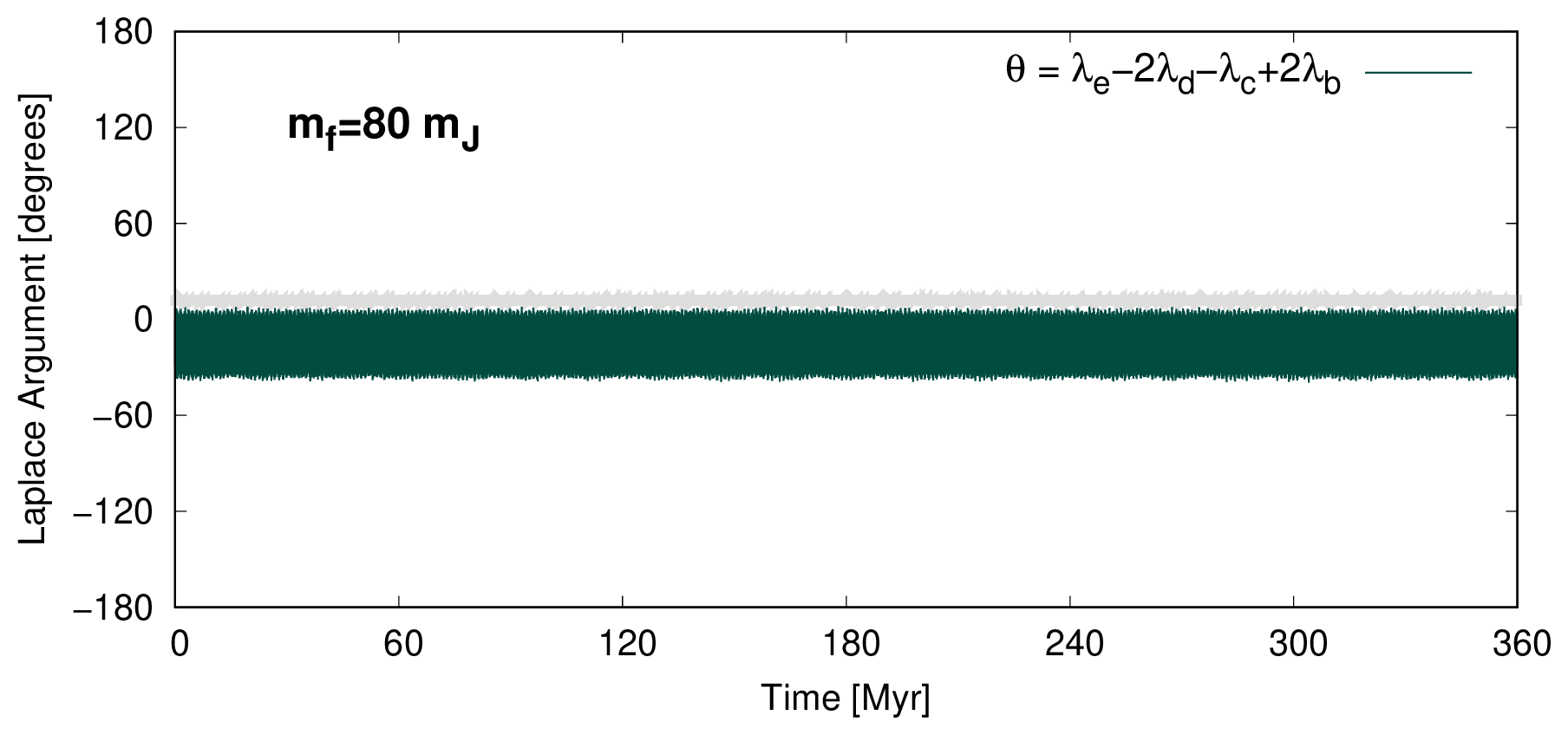}}
}
\centerline{
\hbox{\includegraphics[width=0.5\textwidth]{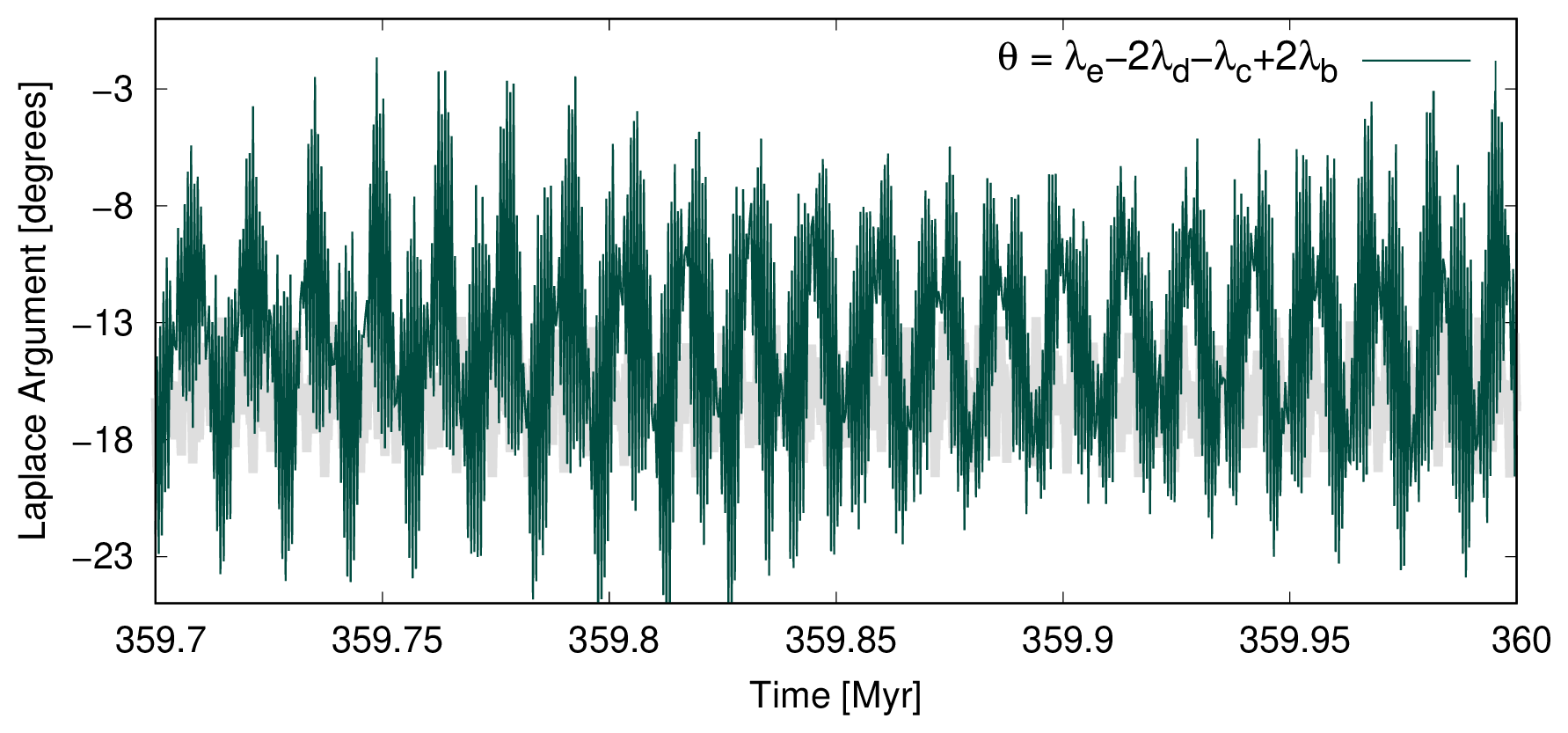}}
\hbox{\includegraphics[width=0.5\textwidth]{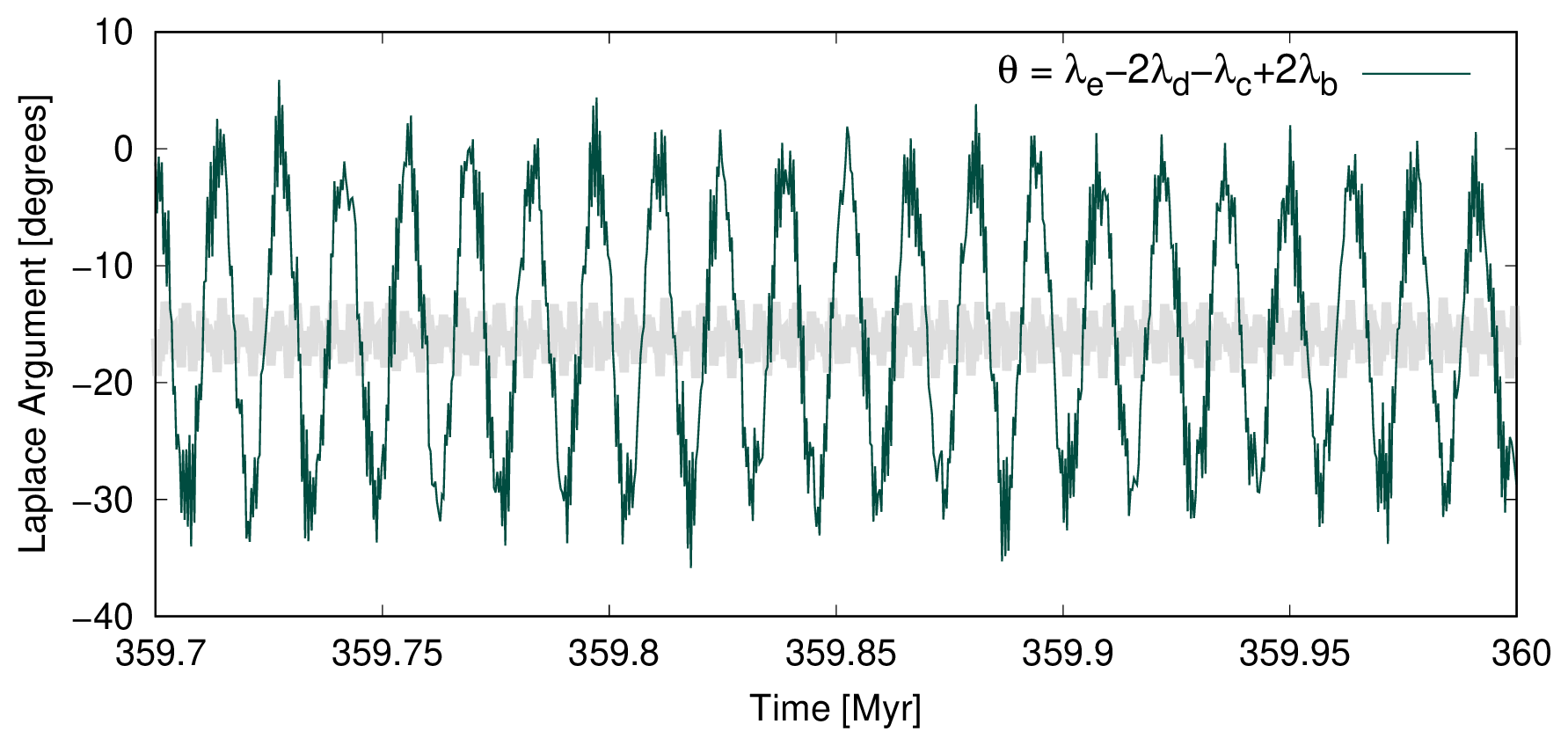}}
}
}
\caption{ 
{\bf Orbital evolution for Models~40 (Left) and~80 (Right) for solutions in the secondary GaiaPMEX solutions with semi-major-axis below 1\au.} 
{\em Top row}: Evolution of the planets semi-major axes over 360 Myr,  integrated with the symplectic integrator SABA4  \citep{Laskar2001} with step size of 64~days. 
{\em Middle row}: Eccentricity evolution during the last 300\,Kyr of the integration interval.
{\em Bottom row}: Evolution of the Laplace argument
$\theta_{8:4:2:1} = \lambda_{\rm e} - 2 \lambda_{\rm d}- \lambda_{\rm c}+2\lambda_{\rm b}$ at the end of the integration interval. 
In the semi-major-axis and eccentricity plots, the background shaded graphs show the variations of the elements in the exact Laplace resonance (periodic orbit) for Model~1 in \citep{Zurlo2022}. 
The orbital elements are plotted as geometric elements derived in the Jacobi frame to suppress spurious variability present in the astrocentric frame. See \citep{Zurlo2022} for details. 
}
\label{fig:EX-els1444}
\end{figure*}

 \clearpage

\begin{figure*}
\centerline{
\vbox{
\hbox{\includegraphics[width=0.9\textwidth]{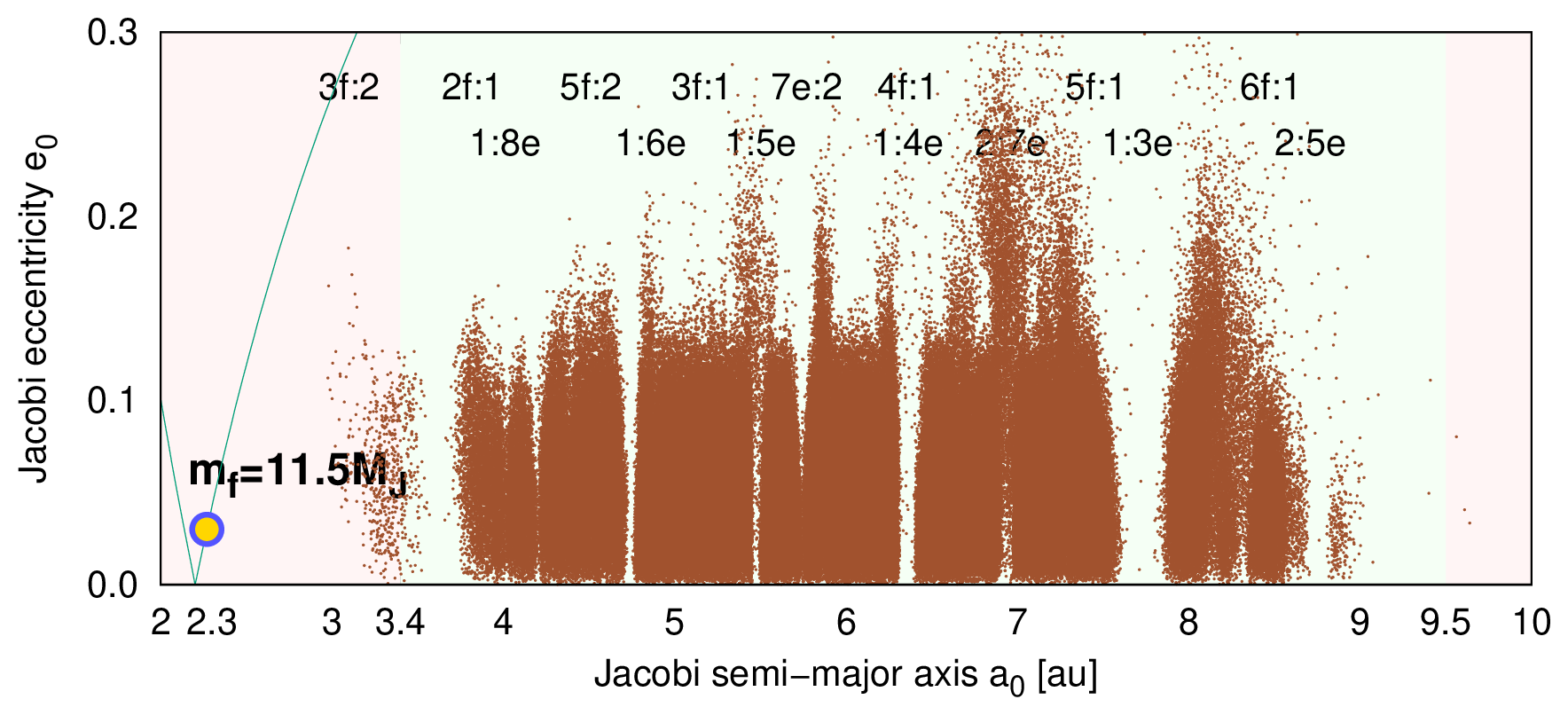}}
\hbox{\includegraphics[width=0.9\textwidth]{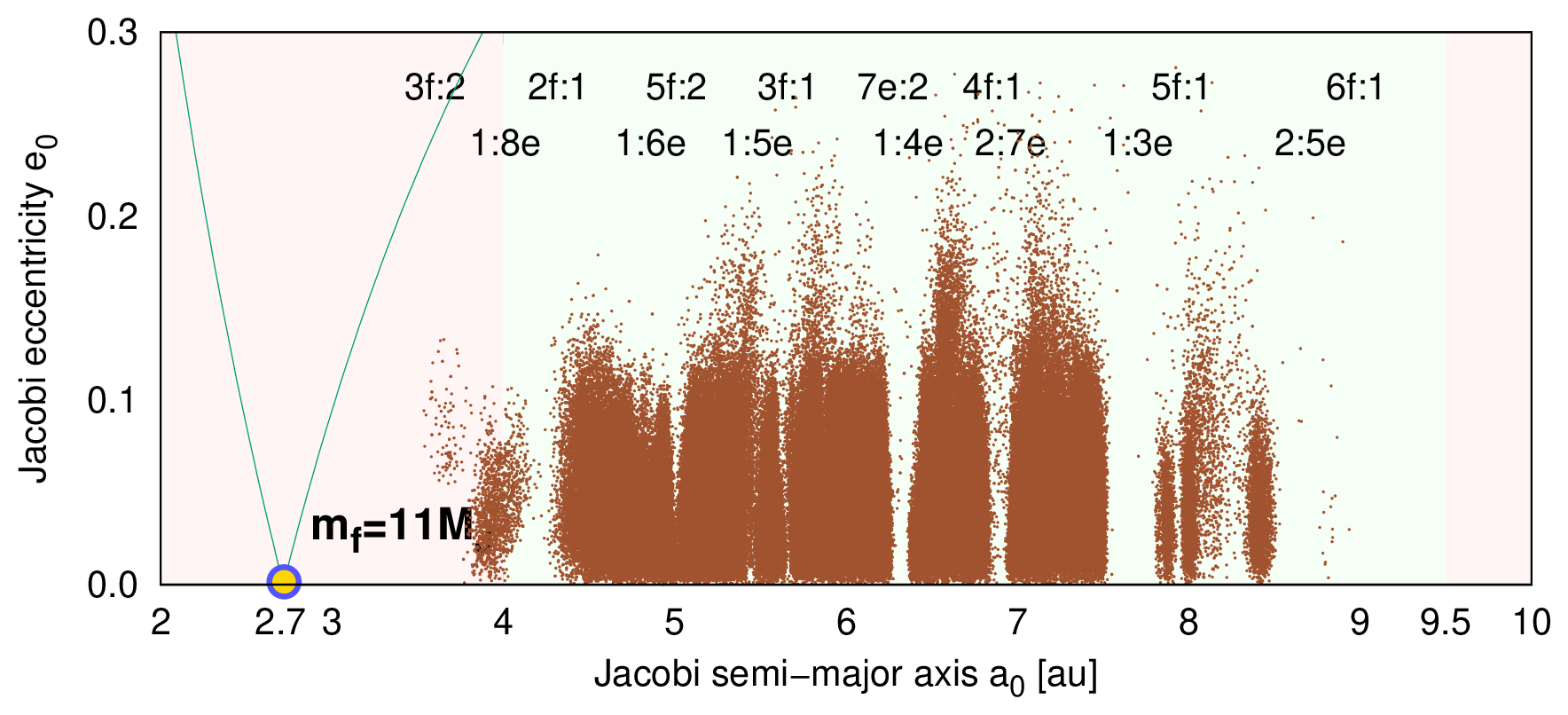}}
\hbox{\includegraphics[width=0.9\textwidth]{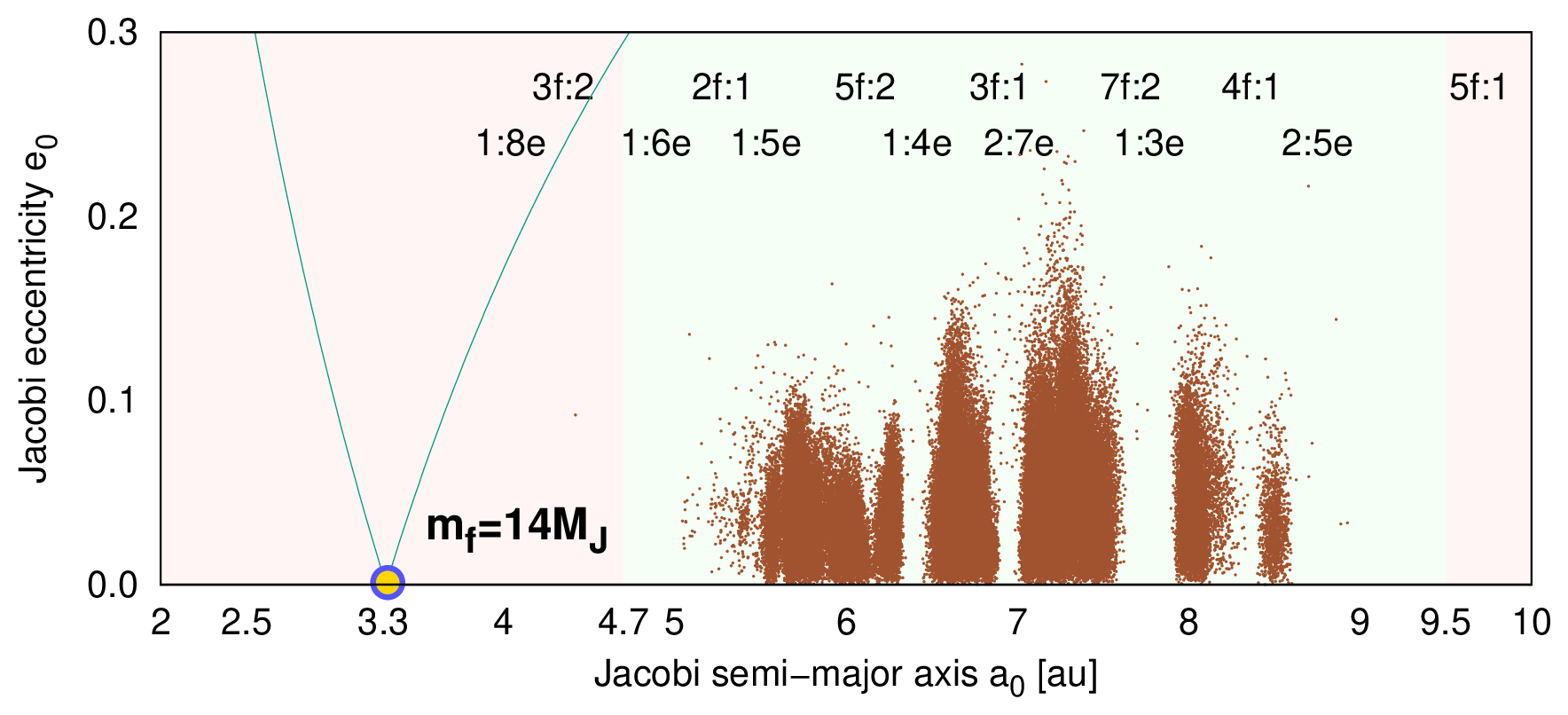}}
}
}
\caption{ 
The orbital status of inner warm disk particles in sma-eccentricity plane expressed in Jacobian reference frame for $m_{\rm f}$ =11.5, 11, and 14\mJup, in Models 115, 11, and 14, respectively, in \edtab{tab:EX-models}. In Model~115, we added a small eccentricity $e_{\rm f}=0.03$ within the extended stability zone (Fig.~\ref{fig:dynmaps}).  Thin curves meeting in the position of planet~f represent the orbital collision curves of the orbits. Empty strips of different widths are Kirkwood-like gaps in the Solar System, due to disk clearing by overlapping of the low-order MMRs with planet~f and planet~e, respectively. They are labeled in the upper part of the plots. Shaded regions are bordered by critical semi-major axes found with the Hill criterion, see SI. 
}
\label{fig:EX-diskselements}
\end{figure*}
\newpage
\clearpage

\subsection*{Extended Data Tables}
{\singlespacing
\begin{longtable}[H]{ccccccccc}
\caption{
\textbf{HARPS RV data for HR 8799}. See \citep{Lagrange2013} for  descriptions of the data and of the data reduction.
}
\label{tab:EX-RV} 
\\
\small$\mbox{JDB}-2454000$ &	
\small$V_{\rm rad}$ & 
\small$\sigma\,V_{\rm rad}$ &
\small$\mbox{JDB}-2454000$ &	
\small$V_{\rm rad}$ & 
\small$\sigma\,V_{\rm rad}$ &
\small$\mbox{JDB}-2454000$ &
\small$V_{\rm rad}$ & 
\small$\sigma\,V_{\rm rad}$ \\
\small$\mbox{[days]}$ &	
\small$\mbox{[km/s]}$ & 
\small$\mbox{[km/s]}$ &
\small$\mbox{[days]}$ &	
\small$\mbox{[km/s]}$ & 
\small$\mbox{[km/s]}$ &
\small$\mbox{[days]}$ &	
\small$\mbox{[km/s]}$ & 
\small$\mbox{[km/s]}$ \\
\hline\small
 986.894 & -1.1104 & 0.0439 & 2794.913 & -0.0837 & 0.0389 & 3334.506 &  0.0096 & 0.0455 \\
 986.909 & -1.2930 & 0.0493 & 2794.915 & -0.1905 & 0.0381 & 3334.508 & -0.0018 & 0.0512 \\
 986.918 & -1.3811 & 0.0370 & 2794.918 & -0.1609 & 0.0382 & 3334.510 & -0.0050 & 0.0462 \\
 986.924 & -1.4224 & 0.0383 & 2794.920 & -0.0598 & 0.0355 & 3334.512 & -0.1276 & 0.0483 \\
 987.922 & -0.7525 & 0.0203 & 2794.923 & -0.1662 & 0.0358 & 3334.514 & -0.1153 & 0.0455 \\
1131.571 &  1.0442 & 0.0224 & 2794.925 & -0.0796 & 0.0365 & 3334.515 & -0.1106 & 0.0408 \\
1131.576 &  0.9817 & 0.0214 & 2794.928 & -0.0781 & 0.0368 & 3334.517 & -0.1123 & 0.0416 \\
1131.583 &  0.8985 & 0.0180 & 2795.915 &  0.5860 & 0.0369 & 3334.519 & -0.2146 & 0.0435 \\
1131.589 &  0.8163 & 0.0175 & 2795.918 &  0.6112 & 0.0333 & 3334.520 & -0.2251 & 0.0455 \\
1131.595 &  0.7556 & 0.0174 & 2795.920 &  0.6385 & 0.0328 & 3334.522 & -0.2699 & 0.0461 \\
1131.600 &  0.6894 & 0.0200 & 2795.923 &  0.7664 & 0.0352 & 3334.524 & -0.3180 & 0.0507 \\
1348.855 & -1.3085 & 0.0294 & 2795.925 &  0.7225 & 0.0346 & 3334.527 & -0.2484 & 0.0525 \\
1348.861 & -1.3359 & 0.0347 & 2795.927 &  0.8343 & 0.0339 & 3334.529 & -0.3013 & 0.0531 \\
1348.867 & -1.2387 & 0.0331 & 2795.930 &  0.9029 & 0.0346 & 3334.533 & -0.3457 & 0.0463 \\
1348.873 & -1.3293 & 0.0272 & 2795.932 &  0.8972 & 0.0341 & 3334.535 & -0.3217 & 0.0453 \\
1348.879 & -1.3450 & 0.0261 & 2795.935 &  0.9639 & 0.0325 & 3334.536 & -0.3486 & 0.0449 \\
1348.885 & -1.3669 & 0.0228 & 2796.913 & -1.0024 & 0.0400 & 3334.540 & -0.3954 & 0.0404 \\
1348.891 & -1.3964 & 0.0204 & 2796.919 & -0.6541 & 0.0389 & 3595.782 &  2.1223 & 0.0432 \\
1348.897 & -1.3567 & 0.0196 & 2796.922 & -0.5854 & 0.0396 & 3595.785 &  2.1017 & 0.0404 \\
1348.903 & -1.4122 & 0.0196 & 2796.925 & -0.5819 & 0.0384 & 3595.789 &  2.1938 & 0.0409 \\
1348.909 & -1.3201 & 0.0195 & 2796.928 & -0.4359 & 0.0369 & 3595.792 &  2.1368 & 0.0418 \\
2583.589 &  0.8544 & 0.0364 & 2796.932 & -0.3358 & 0.0374 & 3595.795 &  2.1416 & 0.0432 \\
2583.592 &  0.8406 & 0.0363 & 2797.910 & -1.0533 & 0.0545 & 3595.798 &  2.1913 & 0.0408 \\
2583.594 &  0.8012 & 0.0398 & 2797.913 & -1.0637 & 0.0468 & 3595.802 &  2.1011 & 0.0388 \\
2583.597 &  0.7637 & 0.0367 & 2797.915 & -1.0549 & 0.0478 & 3595.805 &  2.1587 & 0.0370 \\
2583.599 &  0.7439 & 0.0348 & 2797.918 & -1.0948 & 0.0482 & 3595.808 &  2.1272 & 0.0353 \\
2583.601 &  0.7489 & 0.0359 & 2797.923 & -0.9655 & 0.0534 & 3595.811 &  2.0980 & 0.0321 \\
2583.604 &  0.6103 & 0.0389 & 2797.928 & -1.1690 & 0.0540 & 3713.505 & -0.0814 & 0.0543 \\
2583.606 &  0.6446 & 0.0365 & 2797.935 & -1.0995 & 0.0504 & 3713.507 & -0.0805 & 0.0516 \\
2583.609 &  0.5797 & 0.0395 & 2797.937 & -1.0979 & 0.0525 & 3713.508 &  0.0293 & 0.0517 \\
2583.611 &  0.5196 & 0.0341 & 2798.912 & -0.5692 & 0.0345 & 3713.511 & -0.0146 & 0.0459 \\
2583.614 &  0.5313 & 0.0344 & 2798.918 & -0.4932 & 0.0309 & 3713.513 & -0.0262 & 0.0441 \\
2583.616 &  0.4720 & 0.0338 & 2798.924 & -0.5725 & 0.0333 & 3713.514 &  0.0022 & 0.0422 \\
2583.619 &  0.4050 & 0.0324 & 2798.930 & -0.5270 & 0.0320 & 3713.516 & -0.0448 & 0.0401 \\
2583.621 &  0.3952 & 0.0329 & 2798.936 & -0.4697 & 0.0304 & 3713.517 &  0.0400 & 0.0383 \\
2583.624 &  0.3653 & 0.0316 & 2844.846 & -0.6509 & 0.0534 & 3713.519 & -0.0107 & 0.0393 \\
2583.626 &  0.3209 & 0.0312 & 2844.848 & -0.7001 & 0.0529 & 3713.520 &  0.0258 & 0.0369 \\
2583.628 &  0.2959 & 0.0348 & 2844.859 & -0.7880 & 0.0508 & 3713.522 & -0.0355 & 0.0355 \\
2583.631 &  0.2405 & 0.0420 & 2844.863 & -0.8102 & 0.0528 & 3713.523 &  0.0295 & 0.0386 \\
2583.633 &  0.1824 & 0.0328 & 2844.867 & -0.8730 & 0.0514 & 3713.525 &  0.0210 & 0.0380 \\
2583.636 &  0.1816 & 0.0323 & 2844.876 & -0.8279 & 0.0521 &  &  &  \\
2794.910 & -0.1951 & 0.0346 & 2844.886 & -0.7259 & 0.0511 &  &  &  \\
\hline
\end{longtable}
}
\newpage
\clearpage

\begin{table*}[h]
\centering
\caption{\textbf{ Priors for the MCMC fit of Gaia data, taking into account RV and HCI detection limits. "sma" stands for semi-major-axis.}
}
\label{tab:EX-MCMCpriors}
\begin{tabular}{ccc}
\hline

sma $a$& [0.1, 10]\,au &log uniform \\

mass $m_{\rm f}$ & [0.1, 100]\mJup &log uniform \\

eccentricity $e$ & [0.001, 0.2] &uniform \\

inclination $I$ & $26^{\circ}.873$ & \\

node $\Omega$ & $62^{\circ}.1852189$ & \\

pericenter $\omega$ & [$0^{\circ}$, $360^{\circ}$] & uniform \\

phase & [0, 1] & uniform \\

age & [0.012, 0.023] Gyr & uniform \\

$m_1$ & [1.23, 1.71]\mSun & ${\cal N}$(1.47, 0.12) \\

parallax $\Pi$ &[24.3, 24.6]\,mas & ${\cal N}$( 24.4620, 0.0455 )  \\

offset $v_0$ & [-1, 1] km/s& uniform \\

jitter & [1.5, 3] km/s & uniform \\
\hline
\end{tabular}
\end{table*}

\newpage
 \clearpage
\begin{table*}[h]
\centering
\caption{\textbf{Osculating orbits of planets bcde at 2016.}Keplerian parameters of the osculating orbits for the planets bcde propagated from Model~1 of \citep{Zurlo2022} at epoch 1998.83 to epoch 2016.0.
}
\label{tab:EX-Model1}
\begin{tabular}{lccccccc}
\hline\hline
\multicolumn{8}{c}{\textbf{Keplerian model -- initial epoch 1998.83}} \\
\hline
Planet &
$m\,[M_{\rm Jup}]$ &
$a\,[{\rm au}]$ &
$e$ &
$i\,[^\circ]$ &
$\Omega\,[^\circ]$ &
$\omega\,[^\circ]$ &
$M\,[^\circ]$ \\
\hline
e & 7.40609 & 16.10389 & 0.14768 & 26.873 & 62.185 & 110.817 & 336.828 \\
d & 9.18999 & 26.45173 & 0.11460 & 26.873 & 62.185 &  29.059 &  60.175 \\
c & 7.79612 & 40.98618 & 0.05381 & 26.873 & 62.185 &  92.388 & 145.535 \\
b & 5.80236 & 71.14279 & 0.01667 & 26.873 & 62.185 &  42.970 & 310.895 \\
\hline
\multicolumn{8}{c}{\textbf{Keplerian model -- initial epoch 2015.83}} \\
\hline
Planet &
$m\,[M_{\rm Jup}]$ &
$a\,[{\rm au}]$ &
$e$ &
$i\,[^\circ]$ &
$\Omega\,[^\circ]$ &
$\omega\,[^\circ]$ &
$M\,[^\circ]$ \\
\hline
e & 7.40609 & 16.01179 & 0.14555 & 26.873 & 62.185 & 105.968 &  96.372 \\
d & 9.18999 & 26.14907 & 0.09270 & 26.873 & 62.185 &  24.117 & 123.113 \\
c & 7.79612 & 42.60408 & 0.02557 & 26.873 & 62.185 & 133.005 & 131.936 \\
b & 5.80236 & 69.20613 & 0.01729 & 26.873 & 62.185 & 207.008 & 157.180 \\
\hline\hline

\end{tabular}
\end{table*}
\newpage
\clearpage

{\large 
\begin{table*}[ht]
\centering
\caption{
{\bf Orbital osculating elements for stable, resonant 5-planet configurations.}
The elements are Jacobian, relative to the Laplace plane. 
The stellar mass is $1.47\,\mathrm{M}_{\odot}$. Various masses for planet~f are considered: $10, 11, 11.5, 14, 18, 40$ and $80$\mJup (Models 10-80). Angles are expressed in radians, inclination and nodal angles are $0$\,degrees. We quote 7~decimal places to allow an exact reproduction of the results of the numerical integrations and dynamical REM maps. The initial conditions are for an arbitrary epoch, at which $a_{\rm f} \simeq 16 \au$ in Model 1 of Zurlo (2022), i.e., when the orbits of the bcde planets in these reference systems closely match its resonant configuration with all characteristics explained in SI.
}
\label{tab:EX-models}
\begin{tabular}{@{}lrrrrr@{}}
\toprule
Model & $m$ [$M_{\mathrm{Jup}}$] & $a$ [au] & $e$ & $\omega$ & $\mathcal{M}$ \\
\midrule
& 10.00 &  3.0885756 & 0.0003887 & 5.7809493 & 5.2180579 \\
&  7.41 & 16.0416713 & 0.1295599 & 5.3829344 & 0.7691292 \\
\textbf{Model~10} 
&  9.19 & 25.6852462 & 0.1052987 & 3.9364041 & 4.8524084 \\
&  7.80 & 42.4208626 & 0.0360010 & 2.2607138 & 0.7899586 \\
&  5.80 & 70.8284902 & 0.0090593 & 5.2904446 & 3.7036741 \\
\midrule
& 11.00 & 2.7206913 & 0.0015004 & 3.0135661 & 4.4988919 \\
& 7.41 & 16.1096680 & 0.1409977 & 2.8302302 & 0.0696959 \\
\textbf{Model~11}
& 9.19 & 25.6930133 & 0.1124224 & 4.5133936 & 4.7755206 \\
& 7.80 & 42.4664685 & 0.0357489 & 2.8802829 & 0.7823394 \\
& 5.80 & 71.1067966 & 0.0053178 & 0.2877327 & 6.1454185 \\
\midrule
& 11.46 &  2.2780513 & 0.0003943 &  4.7577000 & 1.7563388 \\
&  7.41 & 16.1472956 & 0.1440553 &  6.0081374 & 1.7007138 \\ 
\textbf{Model~115} 
&  9.19 & 25.8105781 & 0.1072888 &  1.1992685 & 2.6577972 \\ 
&  7.80 & 42.2803857 & 0.0308750 &  6.0556068 & 2.7763197 \\ 
&  5.80 & 71.1681704 & 0.0083288 &  2.0818369 & 5.3896914 \\
%
\midrule
& 14.00 &  3.3254907 & 0.0011222 & 4.2267350 & 3.2587567 \\
&  7.41 & 16.1527467 & 0.1429079 & 4.6977150 & 0.8172382 \\
\textbf{Model~14} 
&  9.19 & 26.0489321 & 0.1031157 & 6.2006705 & 5.3020819 \\
&  7.80 & 42.5509661 & 0.0339009 & 4.3937362 & 1.2756407 \\
&  5.80 & 71.6990868 & 0.0066060 & 0.2309529 & 4.8864127 \\
\midrule
& 18.00 &  3.1681943 & 0.0006506 & 4.8753266 & 5.2670112 \\
&  7.41 & 16.1433497 & 0.1357468 & 5.3962875 & 0.3582453 \\
\textbf{Model~18} 
&  9.19 & 25.5549170 & 0.1035681 & 3.7826577 & 1.6325841 \\
&  7.80 & 42.4177365 & 0.0365410 & 5.3042007 & 5.7120606 \\
&  5.80 & 70.9726321 & 0.0067672 & 2.1919092 & 5.8380444 \\
\midrule
& 40.00 &  0.9203293 & 0.0000487 & 6.0084044 & 6.1010789 \\
&  7.41 & 16.0820731 & 0.1462759 & 0.5503045 & 0.7717953 \\   
\textbf{Model~40} 
&  9.19 & 26.0748246 & 0.1085391 & 2.1990555 & 5.2001637 \\   
&  7.80 & 42.2261789 & 0.0314913 & 0.8742510 & 3.8747404 \\   
&  5.80 & 71.4477215 & 0.0067575 & 4.7730396 & 1.0718639 \\ 
\midrule
& 80.00 &  0.6528589 & 0.0000223 & 5.3842825 & 1.9072589 \\
&  7.41 & 16.0301408 & 0.1457256 & 2.7688320 & 4.8841358 \\  
\textbf{Model~80} 
&  9.19 & 25.8168450 & 0.0978085 & 4.1866256 & 1.1235336 \\  
&  7.80 & 42.1383947 & 0.0276518 & 2.4836759 & 2.2363341 \\  
&  5.80 & 71.1533183 & 0.0105739 & 5.2332843 & 4.7429599 \\ 
\bottomrule
\end{tabular}
\end{table*}
}
\newpage
 \clearpage


\section*{\centerline{Supplementary Information}}

\subsection*{S1. Dynamical modelling of the five-planet system}

\subsubsection*{S1.1 MCOA: adiabatic migration and sampling strategy}

To drive the \host{}bcde system with the added inner planet~f into resonant capture and then extract dissipation-free, present-day observed system configuration, we integrate the full five-body system with \texttt{REBOUND} \citep{Rein2012,Rein2015} and apply dissipative migration forces to a linearly expanded version of the configuration using \texttt{REBOUNDx} \citep{Tamayo2020}, via its \texttt{modify\_orbits\_forces} module. Planets are initialized with astrocentric coordinates, small eccentricities ($e_i=0.01$ for all planets), and random orbital phases (argument of periastron and mean anomaly drawn uniformly). The dissipative forcing follows the standard exponential damping prescription for orbital elements:
\begin{equation}
\dot a_i=-\frac{a_i}{\tau_{a,i}},\qquad \dot e_i=-\frac{e_i}{\tau_{e,i}},
\end{equation}
with $e$-folding time-scales $\tau_{a,i}$, $\tau_{e,i}$, $(i={\rm f}, \ldots, {\rm b})$, implemented as weak non-conservative accelerations.

Because the \host{} system is strongly chaotic and rapidly destabilizes away from resonance, and because migration-driven capture is inherently stochastic, we run an ensemble of simulations to recover resonant configurations consistent with the observed architecture. 
For each inner-companion mass $m_{\rm f}$ within the dominant GaiaPMEX posterior mode and range ($\simeq 10$--$28$\mJup), we initialize the semi-major-axis of planet~f with $a_{\rm f} \in [2.3, 3.3]$\,au (also within $1\sigma$ of the GaiaPMEX constraint). The outer planets are initialized with the semi-major axes and fixed masses of the baseline \citep{Zurlo2022} solution (their Model~1), which we adopt as the reference architecture.

Next, we linearly expand the system by setting  $a_i = \alpha_i\,a_{i,0},$ where $a_{i,0}$ are the reference semi-major axes  and $\alpha_i = [20,21,22,23,24]$ for planets f, e, d, c, and b  respectively.  
This scaling initializes the system in a~dynamically similar, but substantially wider, configuration.

We adopt slow (adiabatic) migration by choosing $\tau_{a,i}$ and $\tau_{e,i}$ in the fiducial runs,
\begin{equation}
\tau_{a,i} = \kappa_i\,\tau,\qquad \kappa_i=[2.1, 2.0, 1.9, 1.8, 1.7], 
\qquad \tau \in[7,8]\,{\rm Myr},
\end{equation}
and uniform $\tau_{e,i} = \tau$, from the innermost to the outermost planet, respectively. The ratio of the $e$-folding times, $\kappa=\tau_{e,i}/\tau_{a,i} \simeq 2$ is very different that expected from physical migration, and serves as a numerical control parameter of the process. This choice enforces convergent inward migration while keeping the orbital evolution slow compared to resonant libration timescales. 

We performed the simulations with  timescales varied by $\pm10\%$ around the fiducial values, to introduce additional variability into the adiabatic  process. We found that the capture outcomes and stability islands are not qualitatively altered.

Each run is integrated with $t_{\rm max}=50$\,Myr, corresponding to several $e$-folding times of the adopted semi-major-axis and eccentricity damping. We sample the orbital state at $N_{\rm orb}=16384$ equally spaced output times. 

Candidate solutions are extracted once the outer architecture reaches the target system linear scale quantified by $a_{\rm e}\simeq 16$\,au, consistent with \citep{Zurlo2022}) in terms of Jacobi elements and  the resonant angles satisfy the capture criteria (S4). We then switch off dissipation and propagate the resulting osculating elements using conservative $N$-body integrations (S5).

\subsubsection*{S1.2. Resonance diagnostics and outcomes}
We monitor the first-order 2:1 resonant angles in the outer chain,  
\begin{equation}
\theta^{(1)}_{i,j} = 2\lambda_j-\lambda_i-\varpi_{i}, \quad 
\theta^{(2)}_{i,j} = 2\lambda_j-\lambda_i-\varpi_{j}, 
\quad i,j = {\rm e},\ldots,{\rm b},
\end{equation}
for subsequent planet-pairs. Here $\lambda$ is the mean longitude and $\varpi$ the longitude of pericentre, expressed as Jacobi geometric elements. We also monitor the Laplace-type critical angle which defines the 8e:4d:2c:1b MMR chain in \citep{Zurlo2022}:
\begin{equation}
\theta_{8:4:2:1}=\lambda_{\rm e}-2\lambda_{\rm d}-\lambda_{\rm c}+2\lambda_{\rm b},
\end{equation}
with low-amplitude librations limited to 90 degrees about a stable centre for the accepted solutions, following their stability analysis conclusions.
We computed the libration amplitude as half the peak-to-peak range of the critical angle, provided there is no circulation for the final $15\%$  of the integration time. Representative capture tracks are shown in \edfig{fig:EX-migration}.

We performed $\simeq 1000$ simulations for each of several $(m_{\rm f}, a_{\rm f})$ pairs, and tested the number of  resonance outcomes. Among them, we found 20\%-40\% systems with libration of all 2-body critical angles, libration of $\theta^{(1)}_{4,5} = 2\lambda_{\rm b} - \lambda_{\rm c} - \varpi_{\rm c}$ around $180$~degrees, librations of $\theta_{8:4:2:1}$ with amplitude less than 90~degrees, as well as the sequence of eccentricities $\simeq 0.14, 0.11, 0.05, 0.02$ for planets e, d, c, and b, respectively. In  $\simeq 5\%$ cases,  the semi-amplitude of $\theta_{8:4:2:1}$ is as small as 10 degrees. All these characteristics reflect the resonant solutions in \cite{Zurlo2022} explaining the relative astrometry. Dynamically, they reproduce this configuration qualitatively and quantitatively to some small variation. Representative solutions for $m_{\rm f}= 10, 11, 11.46, 14,$ and 18\,\mJup with $a_{\rm f}\sim 2-3$\,au are reported in \edtab{tab:EX-models}.

We applied the same procedure for secondary GaiaPMEX posterior modes (Fig.~\ref{fig:EX_CORNER}), centered around $\simeq$ (0.8\au), 40\mJup)  and $\simeq$ (0.5\au, 80\mJup).   We tested a~few hundreds of solutions for each parameter combination. 
We recovered outer bcde Laplace-type resonant chains with essentially the same characteristics as in the dominant posterior mode for semi-major axes as small as $a_{\rm f}\simeq 0.5$au and masses up to $m_{\rm f}\simeq 80$\mJup. Representative outputs are listed in \edtab{tab:EX-models} as Models~40 and~80. This indicates that the adiabatic tuning is robust across the full ($a_f,m_f$) range of interest.

\subsubsection*{S1.3. Long-term $N$-body stability integrations}
To validate the stability of candidate post-migration configurations, we remove dissipation and integrate the full five-planet system with the 4th-order symplectic SABA4 integrator \citep{Laskar2001}. We integrate each representative model for 360\,Myr, exceeding the nominal system age. The fiducial SABA4 step size is 64\,days for the reported long-term runs; additional tests at shorter (32\,days) and longer (128\,days) step sizes were performed for selected models and do not change stability classification. 

We monitor the relative energy error and angular-momentum conservation throughout the integrations, and confirm that the behavior of the resonant angles is insensitive to reasonable changes in the integration time step. For a subset of models, we also used an independent 6th-order symplectic scheme \citep{Beust2024} as a cross-check for representative models. 

\subsubsection*{S1.4. Dynamical maps and chaos indicator (REM)}
To localize the recovered resonant solutions in the phase space and quantify the extent of the stability islands, we compute dynamical maps in the neighbourhood of each representative configuration. We construct grids in:
(i) the $(a_{\rm f},e_{\rm f})$ plane (the innermost companion), and (ii) the $(a_{\rm e},e_{\rm e})$ plane (outer Laplace MMR chain tracer), while keeping the remaining orbital parameters fixed to the representative solution and their nominal values (\edtab{tab:EX-models}).

%
Chaotic (unstable) and regular (quasi-periodic, stable) orbital evolution is quantified using the Reversibility Error Method (REM) \citep{Panichi2017}, a Maximal Lyapunov Exponent-like, CPU-efficient fast indicator computed by integrating the system forward and then backward with a time-reversible symplectic scheme (constant time-step), and measuring the phase-space mismatch between the recovered and initial states. Following \citep{Panichi2017}, we compute REM in canonical Poincar\'e coordinates, using a leapfrog integrator based on the Stumpff solver \citep{Wisdom2016} with a fifth-order symplectic corrector \citep{Wisdom2006}. 

Each REM grid point is integrated for 8-10\,Myr (up to $2 \times 10^4$ outermost orbital periods of \host{}b), sufficient to separate regular from chaotic evolution in \host{}-like multi-giant systems \citep{Panichi2017}. Because the orbital period hierarchy spans nearly two orders of magnitude, we adopt fixed time steps depending on the scanned region: typically, 16\,days for maps involving small moderate eccentricity orbits (up to $e_{\rm f} \simeq 0.4$), and $a_{\rm f} \simeq 2-3$\,au; 8\,days for $a_{\rm f} \lesssim 2$\,au,  and 32-64\,days for maps dominated by the outer chain with initially 
low-eccentricity orbits, chosen to keep relative energy error at $\sim10^{-9}$ or better over the REM interval. In our REM maps, \edfig{fig:dynmaps}, regular long-lived trajectories (dark colors) yield REM values orders of magnitude smaller than chaotic trajectories (bright colors). The stability islands reported in the main text correspond to the contiguous low-$\log{\rm REM}$ regions surrounding the representative solutions. 

\subsection*{S2. Inner debris disk simulations }
Rigorously stable five-body models in \edtab{tab:EX-models} enable the simulation of the inner debris disk in terms of the restricted problem, i.e., planetesimals   evolve due to the gravitational tug of the planets, but do not exert force on the planets and between each other. 

We examined disk persistence and its resonant structure, by varying the nearly-circular orbit of planet~f, with semi-major-axis $\simeq 2.27, 2.72$, and $3.3$\,au, and masses  $11, 11.5$, and $14$\mJup, respectively (Models~11, 115 and 14),  so consistent with the astrometric solution.  

We drew up to $10^6$  massless particles in an annulus $a_0 = [2.5, 10]$\,au with two distributions, $\propto r^2$ (radial quadratic) and $\propto r$ (radial linear), and small initial eccentricity $e_0 \in [0,0.1]$. These limits are roughly consistent with critical semi-major axes that determine bounded orbital motion,  based on the Hill and Wisdom resonance overlap criteria \citep{Wisdom1980}. The Hill criterion is defined in terms of the Hill radius of a~planet, $r_{\rm Hill} = a  (\mu/3)^{1/3}$
where mass parameter is $\mu = m / M_{\star}$.
Particles within $\alpha \simeq 2.5$--$3.5$ Hill radii of a planet are generally unstable, within boundaries determined through $a_{\rm crit} = a \pm \alpha r_{\rm Hill}$. The Wisdom criterion $a_{\rm crit} = a \pm 1.5 a \mu^{2/7}$ is less restrictive than the Hill criterion with $\alpha=3.5$, used here to define analytically raw inner disk boundaries. 

The whole system was then integrated for 35\,Myr with the  GENGA code \citep{Grimm2014, Grimm2022},  an open-source $N$-body simulation code designed to model planetary system evolution and planet formation.  We used a hybrid integrator combined of fixed second-order symplectic Wisdom-Holman symplectic method (W-H leapfrog)  for general orbits with a variable-step Bulirsch-Stoer-Gragg integrator for close encounters, ensuring energy conservation at the level of $10^{-6}$-$10^{-7}$ for five-planet \host{} models. We set the leapfrog step size of 28 and 32\,days, depending on initial $a_{\rm f} \simeq 2.3$, and $\simeq 3.3$,\au. 

As a result, we find a distribution of planetesimals that survived the integration without ejections from the system of collisions with the planets or the star. experiencing neither ejection from the system nor collisions with the planets or the star.
They are illustrated as a snapshot at the final epoch in the Laplace plane of the system (Fig.~\ref{fig:disk11}) and in the $(a_0,e_0)$-plane of osculating, geometric elements in Jacobi frame (\edfig{fig:EX-diskselements}). We identify gaps in the elements distribution as Kirkwood-like gaps in the Main Belt of the Solar System. They correspond mainly to  low-order MMRs with planets~e, and~f, respectively, as labeled in the plots, in terms of orbital mean motion (frequencies) ratios. The overlap of the MMRs with the two planets leads to strongly chaotic evolution and planetesimals removal in the gap regions.

\end{document}